\documentclass[usenatbib,usegraphicx]{mn2e}
\usepackage{aas_macros}
\usepackage{rotating}
\usepackage{longtable}
\usepackage{times} %% For PDF
\def\gs{\mathrel{\raise0.35ex\hbox{$\scriptstyle >$}\kern-0.6em
\lower0.40ex\hbox{{$\scriptstyle \sim$}}}}
\def\ls{\mathrel{\raise0.35ex\hbox{$\scriptstyle <$}\kern-0.6em
\lower0.40ex\hbox{{$\scriptstyle \sim$}}}}

\def\lsim{~\raise0.3ex\hbox{$<$}\kern-0.75em{\lower0.65ex\hbox{$\sim$}}~}
\def\gsim{~\raise0.3ex\hbox{$>$}\kern-0.75em{\lower0.65ex\hbox{$\sim$}}~}
\newcommand{\mum}{$\mu$m~}

\def\gs{\mathrel{\raise0.35ex\hbox{$\scriptstyle >$}\kern-0.6em \lower0.40ex\hbox{{$\scriptstyle \sim$}}}}
\def\ls{\mathrel{\raise0.35ex\hbox{$\scriptstyle <$}\kern-0.6em \lower0.40ex\hbox{{$\scriptstyle \sim$}}}}
\newcommand{\arcsecs}{\mbox{$^{\prime\prime}$}}

\newcommand{\Msolar}{\mbox{M$_{\odot}\,$}}
\newcommand{\Lsolar}{\mbox{L$_{\odot}\,$}}
\newcommand{\degs}{\mbox{$^{o}$}}

\newcommand{\parcsec}{\mbox{$\stackrel{\prime\prime}{\textstyle .}$}}
\newcommand{\parcmin}{\mbox{$\stackrel{\prime}{\textstyle .}$}}
\newcommand{\mumeol}{$\,\mu$m}

\begin{document}

\title[A 1200\mum MAMBO Survey of the ELAIS\,N2 and Lockman Hole: I]{
      A 1200\mum MAMBO survey of ELAIS\,N2 and the Lockman Hole: I.\ Maps, sources and number counts}

\author[Greve et al.]{
\parbox[t]{\textwidth}{
\vspace{-1.0cm}
T.\ R.\ Greve$^{1}$,
R.\ J.\ Ivison$^{1,2}$,
F.\ Bertoldi$^{3}$,
J.\ A.\ Stevens$^{2}$,
J.\ S.\ Dunlop$^{1}$,
D.\ Lutz$^{4}$ \&
C.\ L.\ Carilli$^{5}$
}
\vspace*{6pt}\\
$^{1}$ Institute for Astronomy, University of Edinburgh, Blackford Hill, Edinburgh EH9 3HJ, UK\\
$^{2}$ Astronomy Technology Centre, Royal Observatory, Blackford Hill, Edinburgh EH9 3HJ, UK\\
$^{3}$ Max-Planck-Institut f\"{u}r Radioastronomie(MPIfR), Auf dem H\"{u}gel 69, 53121 Bonn, Germany\\
$^{4}$ Max-Planck-Institut f\"{u}r extraterrestrische Physik, Postfach 1312, 85741, Garching, Germany\\
$^{5}$ National Radio Astronomy Observatory, P.O.~Box O, Socorro, NM 87801, USA
\vspace*{-0.5cm}}

%\date{\fbox{\sc Draft dated: \today\ }}
%\date{Accepted ... ; Received ... ; in original form ...}

\pagerange{000--000}

\maketitle

\begin{abstract}
We present a deep, new 1200\mum survey of the ELAIS\,N2 and Lockman Hole
fields using the Max Planck Millimeter Bolometer array (MAMBO).  The areas
surveyed are 160 arcmin$^2$ in ELAIS\,N2 and 197 arcmin$^2$ in the Lockman
Hole, covering the entire SCUBA `8\,mJy Survey'. In total, 27 (44) sources
have been detected at a significance $\ge$4.0$\sigma$ ($\ge$3.5$\sigma$). The
primary goals of the survey were to investigate the reliability of (sub)millimetre
galaxy (SMG) samples, to analyse SMGs using flux ratios sensitive to redshift
at $z\rm > 3$, and to search for `SCUBA drop-outs', i.e.~galaxies at $z\rm >>
3$.  We present the 1200\mum number counts and find evidence of a fall at
bright flux levels. Employing parametric models for the evolution of the local
60\mum {\em IRAS} luminosity function (LF), we are able to account
simultaneously for the 1200 and 850\mum counts, suggesting that the MAMBO
and SCUBA sources trace the same underlying population of high-redshift,
dust-enshrouded galaxies. From a nearest-neighbour clustering analysis we find
tentative evidence that the most significant MAMBO sources come in pairs,
typically separated by $\sim$23$''$. Our MAMBO observations unambiguously
confirm around half of the SCUBA sources. In a robust sub-sample of 13 SMGs
detected by both MAMBO and SCUBA at a significance $\ge$3.5$\sigma$, only one
has no radio counterpart. Furthermore, the distribution of 850/1200\mum flux
density ratios for this sub-sample is consistent with the spectroscopic
redshift distribution of radio-detected SMGs (Chapman et al.~2003).
Finally, we have searched for evidence of a high-redshift tail of SMGs
amongst the 18 MAMBO sources which are not detected by SCUBA. While
we cannot rule out that some of them are SCUBA drop-outs at $z>>\rm 3$,
their overall 850-to-1200\mum flux distribution is statistically
indistinguishable from that of the 13 SMGS which were robustly identified 
by both MAMBO and SCUBA.
\end{abstract}


\section{Introduction}

In a time of `high-precision cosmology' one of the fundamental
questions about which we remain largely ignorant is the formation and
evolution of galaxies and clusters of galaxies. One of the most
important breakthroughs in this field was the discovery of a
significant population of far-IR-luminous, high-redshift sources in
surveys at submillimetre (submm) and millimetre (mm) wavelengths using
SCUBA and MAMBO (Smail, Ivison \& Blain 1997; Hughes et al.~1998;
Barger et al.~1999; Eales et al.~2000; Bertoldi et al.~2000),
resolving at least half of the far-IR/submm background detected by the
{\em DIRBE} and {\em FIRAS} experiments (e.g.~Hauser et al.~1998).

It is widely believed that the large far-IR luminosities
($\gs$10$^{12}\,\Lsolar$) of these sources is caused by intense UV
light from starbursts and/or active galactic nuclei (AGN) being
absorbed by dust and re-radiated longwards of $100$\mum. The
negative $k$-correction at $\lambda \ge400$\mum allows
submm/mm observations to select star-forming galaxies at $z > 1$ in an
almost distance-independent manner, providing an efficient method of
finding obscured, star-forming galaxies at 1 $< z <$ 10 (Blain \&
Longair 1993).

The large star-formation rates ($\sim$1000\,$\Msolar\,\mbox{yr}^{-1}$)
found for (sub)mm galaxies (hereafter SMGs) are sufficient to
construct a giant elliptical ($\sim$10$^{11}\,\Msolar$) in less than a
Gyr, providing that the starburst is continuously fueled.  This has
led people to speculate that SMGs could be the progenitors of such
galaxies (e.g.~Dunlop 2001), a scenario which is further
strengthened by the fact that the co-moving number density of SMGs
appears to be consistent with that of today's massive ellipticals
(Scott et al.~2002; Dunne, Eales \& Edmunds 2003). However, the
nature of SMGs, and in particular their relationship with possible
present-day counterparts, is not known, just as their relation to
other high-redshift populations such as Lyman-break galaxies (LBGs)
and extremely red objects (EROs) is not well understood.

Progress has been hampered by the large positional uncertainties of
the SMGs. The relatively large beams of (sub)mm telescopes
(11--14$''$, {\sc fwhm}) makes it impossible to reliably tie an SMG to
an optical or near-IR counterpart, unless additional data at a
complementary wavelength are available. Since a characteristic feature
of both starburst galaxies and AGN is radio emission, deep radio
imaging has proven to be a highly efficient way of accurately
identifying optical/near-IR counterparts to SMGs (Ivison et al.~1998,
2000, 2002; Smail et al.~2000).

Use of the radio-to-submm spectral index as an redshift indicator
(Hughes et al.~1998; Carilli \& Yun 1999, 2000) has shown that the
SMGs lie at high redshift, with an estimated median redshift of $\ge
2$ (Ivison et al.~2002).  Recently, Chapman et al.\ (2003, 2004) have
obtained spectroscopic redshifts for $\sim$90 SMGs and found they span
0.8 $< z <$ 4, with a median of 2.4, although they cautioned that the
distribution might be skewed towards lower redshifts since a
requirement for getting a spectrum was that the SMGs had $\mu$Jy
counterparts in the radio.

In a very deep radio survey of the Lockman Hole and ELAIS\,N2 fields,
Ivison et al.~(2002) found that about one third of SMGs did not have
radio counterparts. One plausible explanation was that some of these
radio-blank SMGs lie at very high redshifts. Such a population of SMGs
--- the so-called `high-redshift tail' --- would ask difficult
questions of popular hierarchical models. The high-redshift SMGs would
have low $S_{850\mu\rm{m}}/S_{1200\mu\rm{m}}$ flux ratios (see Eales
et al.~2003) and would be readily detectable with MAMBO.

Blain, Barnard \& Chapman (2003), amongst others, have pointed out the
limitations of photometric redshift techniques: the redshift is
degenerate with the far-IR luminosity as parametrised by the dust
temperature, $T_{\rm d}$. In principle, however, a comparison between
the $850$, $1200$\mum and 1.4-GHz flux densities allows us, in some
cases at least, to break this degeneracy, assuming that SMGs follow
the radio/far-IR correlation.  For example, an SMG which has a 'warm'
$S_{850\mu\rm{m}}/S_{1200\mu\rm{m}}$ ratio but has no radio
counterpart is likely to be at high redshift. The
$S_{850\mu\rm{m}}/S_{1200\mu\rm{m}}$ redshift estimator is particular
sensitive at $z>\rm 3$ and is currently the most effective way to
assess whether there is a significant high-redshift tail of SMGs.

The primary advantage of a mapping survey, rather than pointed
photometry (on--off) observations of known sources (e.g.~Eales et
al.~2003), is that one obtains an unbiased view of the sky. Data are
not skewed by the choice of targets, by possible errors in
coordinates, or by potentially spurious assumptions about how the sky
is expected to appear. A key goal of our new survey was to determine
unbiased flux densities for the radio-blank SCUBA sources using MAMBO,
and to search for new populations of mm-bright sources, in particular
a sign of a dusty, star-forming $z\rm >> 5$ population --- `SCUBA
drop-outs' --- that would be expected to be below the typical SCUBA
detection threshold at $850$\mum but detectable by MAMBO at
$1200$\mum.

In this paper we present a new, unbiased $1200$\mum survey, using
MAMBO on the IRAM 30m telescope, of the Lockman Hole and ELAIS\,N2,
the two regions observed by the SCUBA 8\,mJy survey (Scott et al.~2002; Fox et al.~2002). 
A plethora of multi-wavelength observations
exist for both fields, including very deep X-ray, optical, near-IR,
mid-IR and radio imaging (Hasinger et al.~2001; Manners et al.~2002; 
Ivison et al.~2002; Almaini et al.~2002). Our observations, data reduction and maps are described in \S2
and 3. The source extraction technique and the source catalogues for
each field are presented in \S4, as are the results of Monte Carlo
simulations to assess completeness, positional accuracy, flux
boosting, etc. In \S5 and 6 we present our measurements of the source
counts and the clustering properties of MAMBO sources. Finally, \S7
describes our joint analysis of the $850$ and $1200$\mum samples, and
implications for ongoing (sub)mm surveys and for the redshift
distribution of SMGs.

Throughout, we have adopted a flat cosmology, with $\Omega_m=0.3$,
$\Omega_\Lambda=0.7$ and $H_0=70$\,km\,${\rm s^{-1}}$\,Mpc$^{-1}$.

\section{Observations}\label{sec:observations}

The survey was carried out with the 37- and 117-channel MPIfR Max Planck
Millimeter Bolometer arrays (MAMBO-I and MAMBO-II; Kreysa et al.~1998) at the
IRAM 30m telescope on Pico Veleta near Granada in Spain. Both MAMBO-I and
MAMBO-II are He$^3$-cooled arrays operating at an effective frequency of
250\,GHz or $1200$\mum with a half-power spectral bandwidth of 80\,GHz. At
$1200$\mumeol, the 30m telescope has an effective beam of 10.7$''$ ({\sc fwhm}).
The arrays are background limited and the performance of the array routinely
gives noise equivalent flux densities (NEFDs) of 30 -- 45\,mJy\,Hz$^{-1/2}$.  The
bolometer feedhorns are arranged in a compact hexagonal pattern each with a
diameter of $2F\lambda$ which ensures an optimal coupling to the incoming
radiation from a point source.  MAMBO-II is amongst the largest (sub)mm bolometer arrays
currently in use.  In combination with the IRAM 30m dish which has a surface
accuracy of 75\,$\mu$m rms, it is the most powerful tool for large blank-field
surveys at (sub)mm wavelengths, and will remain so until the advent of
APEX/LABOCA, JCMT/SCUBA2 and LMT/Bolocam.

Scan-mapping along the azimuthal direction is the only method available at the
30m to map large areas of the sky. The signal from the sky is modulated by
the secondary mirror (the wobbler) which is wobbling in the the scan direction
(azimuth).  The wobbler frequency is 2\,Hz which reflects a compromise between
wanting to eliminate changes in the atmosphere on as short a timescale as
possible and the challenges involved in moving a 2m secondary at this
frequency and keeping it mechanically stable.  We used a standard on-the-fly
MAMBO scan-map, typically $300\arcsec \times 320\arcsec$ in size, scanned at a velocity
of 5$''$\,s$^{-1}$, and with an elevation spacing between each subscan of
8$''$. Hence, a map consists of 41 subscans of 60s each.  This results in a
fully-sampled map over a $300\arcsec \times 320\arcsec$ region in $\sim$43\,min
(this includes $\sim 3$s of overhead per subscan). 
In order to obtain uniform coverage, a regular grid
with a grid-spacing of 2$\arcmin$ was defined across each of the two
fields. Each grid position was observed once which in practice means that, in
the final map, each point on the sky has been observed at least two times. In
principle, this ensures that an rms of $\sim$0.8\,mJy\,beam$^{-1}$ is reached
across most of the field. In practice, however, scans were taken in slightly
different weather conditions which means that the noise is not entirely
uniform.  In order to eliminate any possible systematics and any residual
effects from the double beam profile, wobbler throws in the range $36\arcsecs - 45\arcsecs$
were used. Furthermore, maps were taken with different scan-directions,
just as care was taken to map each grid position once while the field was
rising and once while it was setting. All this served the purpose of minimising
any systematic effects from the atmosphere and/or instrument which may
otherwise had arisen from observing the same grid-position in two consecutive maps
using identical wobbler- and scan-configuration.

\begin{figure*}
\begin{center}
\includegraphics[width=0.9\hsize,angle=0]{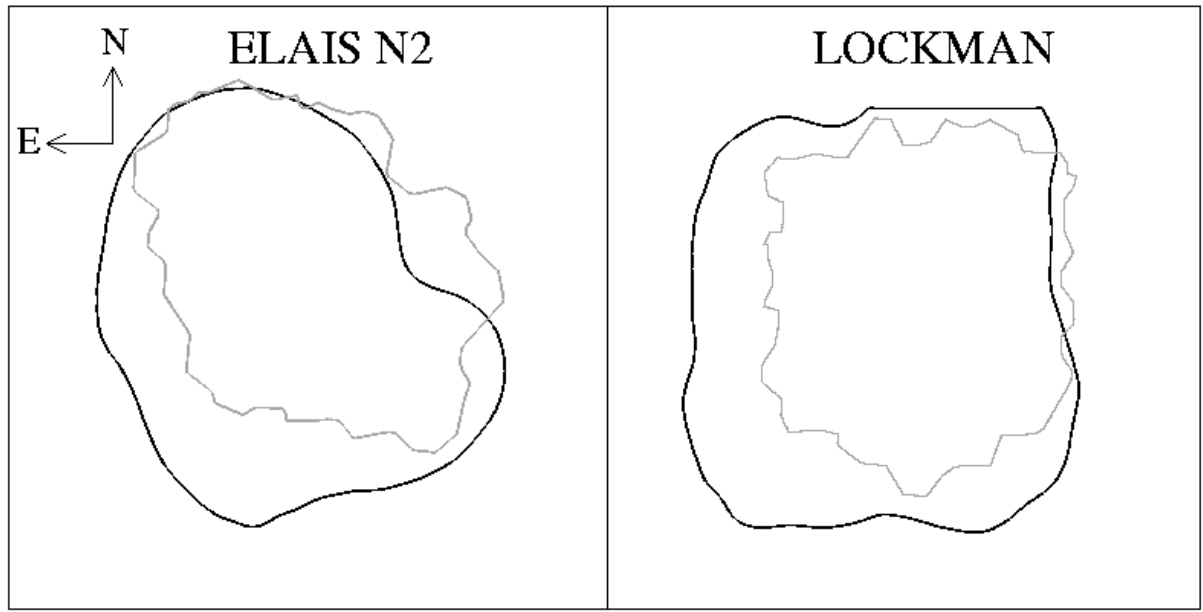} 
\caption[The MAMBO Survey regions.]{The MAMBO survey regions (outlined
in black) of the ELAIS\,N2 and Lockman Hole fields. The mapped areas
are 160 and 197 sq.~arcmin in ELAIS\,N2 and Lockman Hole,
respectively. For comparison, the coverage of the SCUBA UK 8\,mJy
Survey (Scott et al.~2002) is also shown (grey curve). The boxes
are $21\parcmin7\times 21\parcmin7$.}
\label{fig:mambo_scuba_overlay}
\end{center}
\end{figure*}

The two fields were only observed when above $30\degs$ and below $70\degs$ in
elevation. The latter constraint, which avoids distorted scan-maps, was
particularly troublesome for the ELAIS\,N2 field which reaches a maximum
elevation of $85\degs$ at Pico Veleta.

Since observations were pooled, they were done only under good weather
conditions, i.e.~ when the atmospheric zenith opacity at $1200$\mum was less than
0.3, with low sky-background variations. After each map, i.e.~every hour, the
telescope was pointed and focused. The opacity of the atmosphere was measured
every other hour by doing a skydip but was also continuously monitored with a
radiometer located next to the telescope.  Variations in the sky-background
were monitored from the pointings by on-line measurements of the correlation of
the horns across the array. Also quick on-offs of the pointing sources were
done throughout the night in order to check the sky noise and calibration.  In
order to tie down the absolute flux calibration, primary flux calibrators
(including planets when available) were observed at the beginning and end of
each run, and resulted in an absolute flux calibration of better than 20 per
cent.

In total, 34 scan-maps, corresponding to 26\,hrs, went into the final map of
ELAIS\,N2, all of which were obtained with MAMBO-II as part of the pooled
observing mode during the winter period of the 2001-2002 and 2002-2003
seasons. The bulk of the observations of the Lockman Hole were also obtained
during that time using MAMBO-II, although, early observations of the Lockman
Hole during the winter of 2000--01, and to some extent also 2001--02, were
obtained with the 37-channel MAMBO-I array.  For the Lockman Hole, 19 and 51
scan-maps taken with MAMBO-I and MAMBO-II, respectively, were used for the
final map, equivalent to a total integration time of 53\,hr.

The areas surveyed were 160 arcmin$^2$ in the ELAIS\,N2 field and 197
arcmin$^2$ in the Lockman Hole. In the case of ELAIS\,N2 the MAMBO observations
were designed to cover the region observed at $850$\mum as part of the SCUBA UK
8\,mJy Survey (Scott et al.~2002). The small fraction of the SCUBA map which is
not covered by our observations (Fig.~\ref{fig:mambo_scuba_overlay}) is also
the most noisy, and no $850$\mum sources were detected in that region.  However,
in the Lockman Hole the overlap between MAMBO and SCUBA observations is
complete, as is seen from Fig.~\ref{fig:mambo_scuba_overlay}. The Lockman Hole
MAMBO data presented in this paper are part of the larger area MAMBO Deep Field Survey
(Bertoldi et al.~2000; Dannerbauer et al.~2002; Bertoldi et al.~2004, in prep.).
In addition, the Lockman Hole is being targeted by SCUBA as part of the
SCUBA Half Degree Extragalactic Survey (SHADES --- {\it http://www.roe.ac.uk/ifa/shades/}),
and the data presented here constitute only a small part of the area surveyed at (sub)mm wavelengths in
that part of the sky.

\section{Data reduction}

The data were reduced using the {\sc mopsi} software package (Zylka 1998).
Bolometers which were either dead or very noisy were flagged from the data
reduction, and the remaining data streams were de-spiked, flat-fielded and
corrected for atmospheric opacity.

Since all bolometers look through the same region of the atmosphere at any
given time, there will be a correlated signal across the array which will have
to be removed in order to recover the (uncorrelated) astronomical signal. In
{\sc mopsi} the removal of this correlated sky background is done in an
iterative way. For each bolometer an inner and outer radius from the channel is
specified, and the correlation with every bolometer lying within this annulus
is computed. In our case, we chose an inner and outer radius of 1$''$ and
60$''$, respectively, which is suitable for compact weak sources. The
correlated signals of the surrounding 6 channels with the best correlation were
then chosen and the average correlated noise subtracted from the bolometer in
question. This procedure is then repeated until the sky background has been
removed satisfactorily across the array.

Up until this point in the data reduction, the signal from each chopping
position (called the on and off phase) was processed separately, the reason
being that in a single phase the bolometers correlate much better which results
in a much more reliable subtraction of the background. {\sc mopsi} then
calculated the phase difference, thereby effectively removing any electronic
systematics between the data obtained in one wobbler position and the
other. Additional de-spiking and baseline-fitting was then done on the phase
differences, from which the weights for each bolometer were calculated.

Finally, the data were restored and rebinned using a shift-and-add technique
which for each map produces a positive image bracketed by two negative images
of half the intensity located one wobbler throw away. The rebinning was done
onto a grid of 1 square arcsec pixels, with the flux in each pixel being a
noise-weighted average of the bolometers hitting that position.  {\sc mopsi}
also outputs a weight image, $W$, which at pixel position $(i,j)$ is given by
$W(i,j) = \sum_{k} 1/\sigma_{k}^2$, i.e.~the error on the noise-weighted
average, where $k$ denotes summing over the bolometers 'seen' by pixel $(i,j)$.

The noise maps are shown in Figure \ref{fig:noise-maps}. The rms noise in the
deepest parts of the maps is $\sim$0.6\,mJy\,beam$^{-1}$, increasing towards
the edges. The Lockman Hole received more integration time than the ELAIS\,N2
field and as a result it is slightly deeper and has more uniform noise
properties than the latter.

\section{Source extraction and Monte Carlo Simulations}

\subsection{Source extraction}

\begin{figure*}
\begin{center}
\includegraphics[width=0.45\hsize,angle=-90]{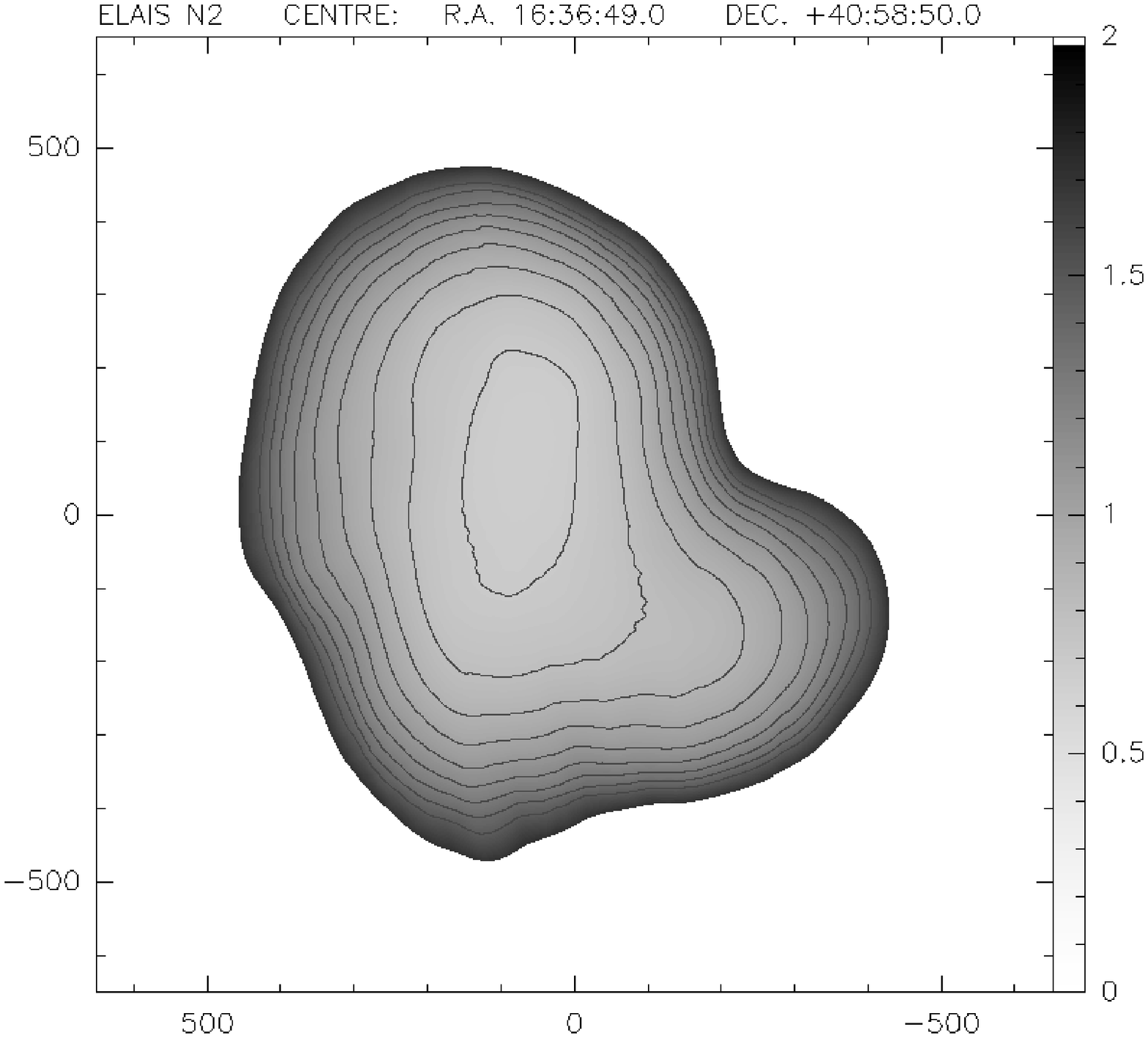} 
\includegraphics[width=0.45\hsize,angle=-90]{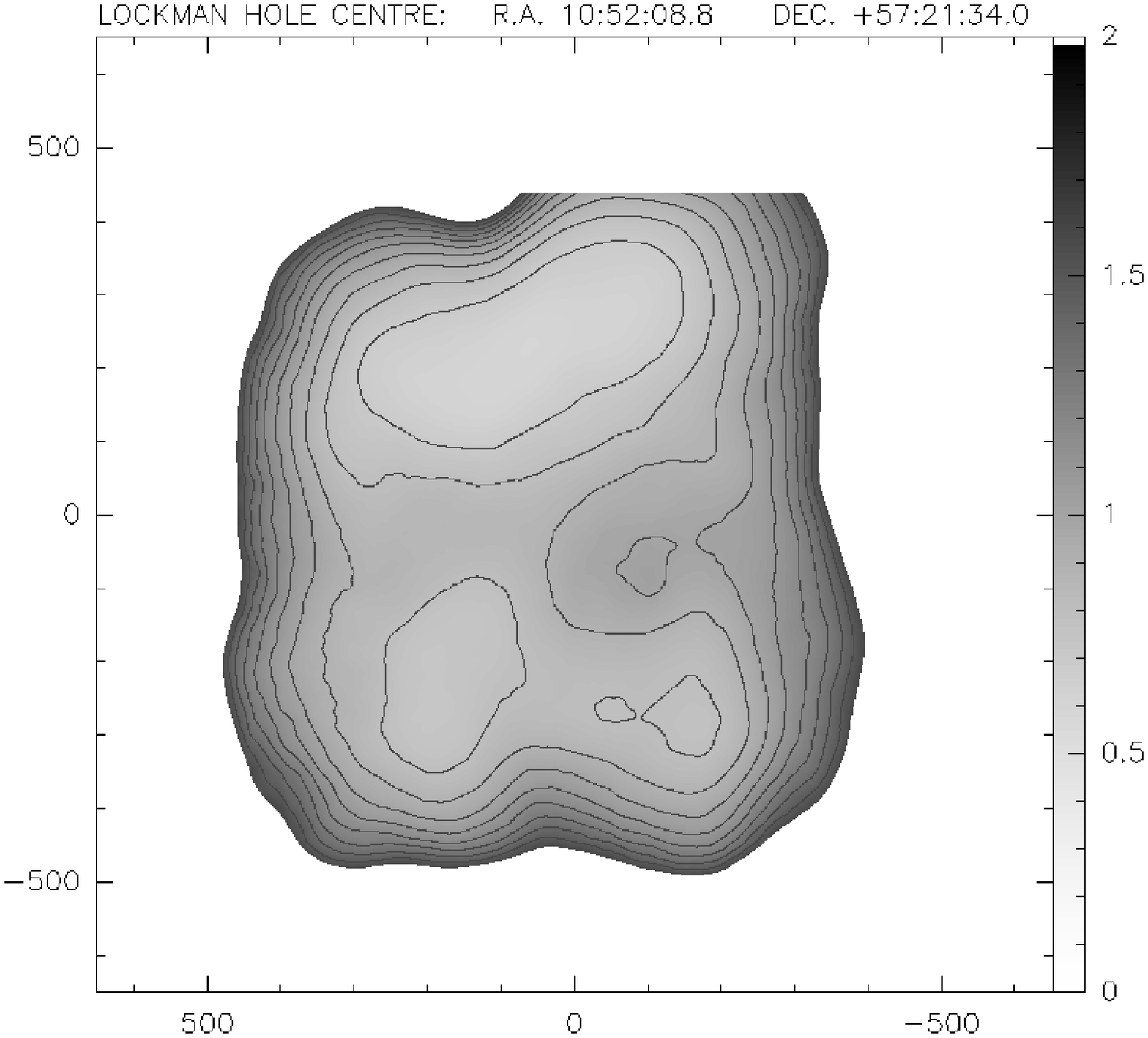} 
\caption[Noise maps of the ELAIS\,N2 and Lockman Hole fields.]  {Noise maps of
the ELAIS\,N2 and Lockman Hole East fields. Contours are at 0.7, 0.8
... 1.5\,mJy\,beam$^{-1}$.  The axes denote the offset (arcsec) from the map
centre, and the intensity scale is in units of mJy\,beam$^{-1}$.  }
\label{fig:noise-maps}
\end{center}
\end{figure*}

\begin{figure*}
\begin{center}
\includegraphics[width=0.9\hsize,angle=-90]{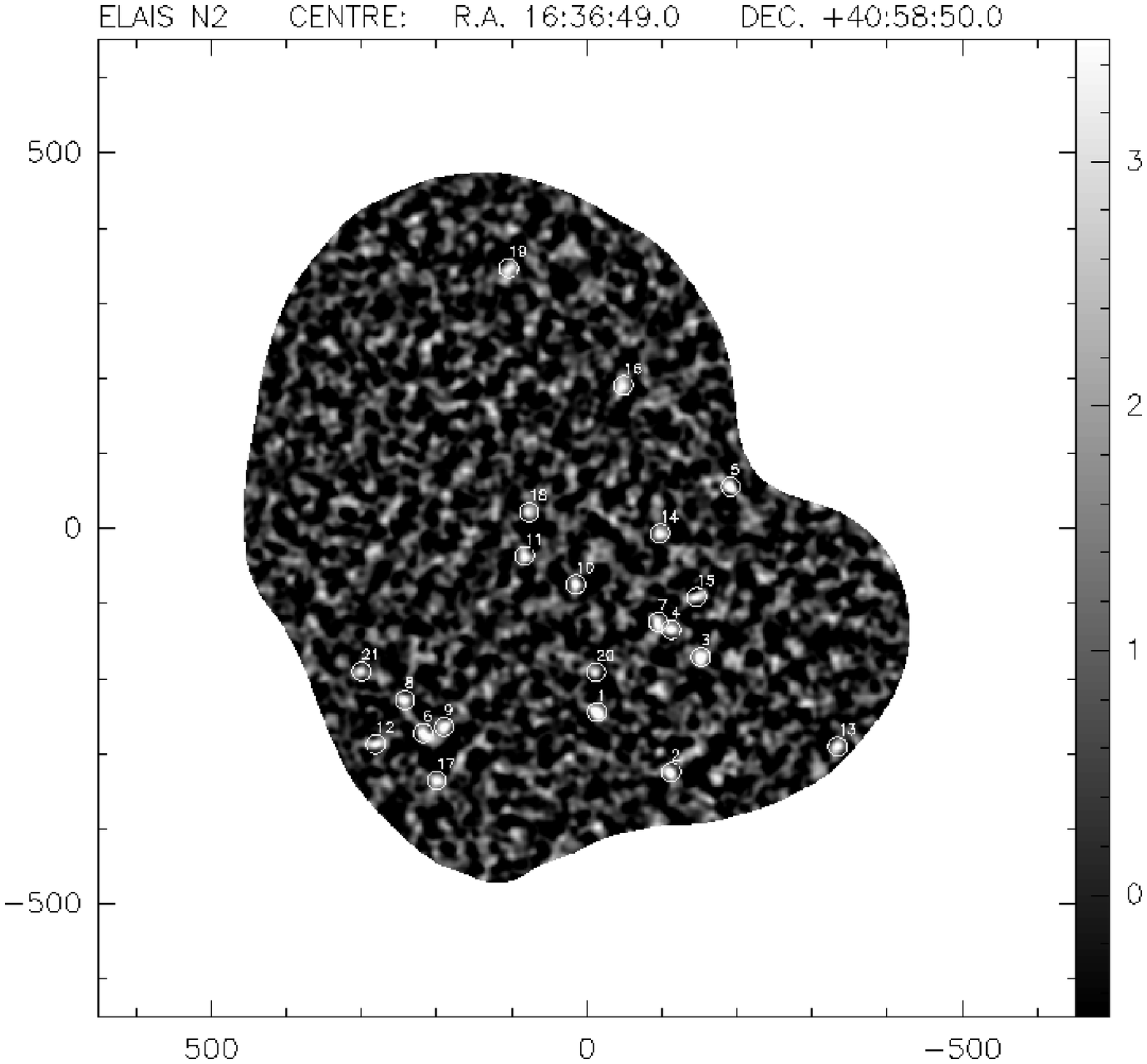} 
\caption[$1200$\mum MAMBO signal-to-noise map of the ELAIS\,N2 field.]{$1200$\mum
MAMBO signal-to-noise map of the ELAIS\,N2 field. Sources detected at
$\ge$$3.5\sigma$ are circled in white, and are numbered in order of
significance. The axes denote the offset (arcsec) from the map centre.}
\label{fig:map-n2}
\end{center}
\end{figure*}

\begin{figure*}
\begin{center}
\includegraphics[width=0.9\hsize,angle=-90]{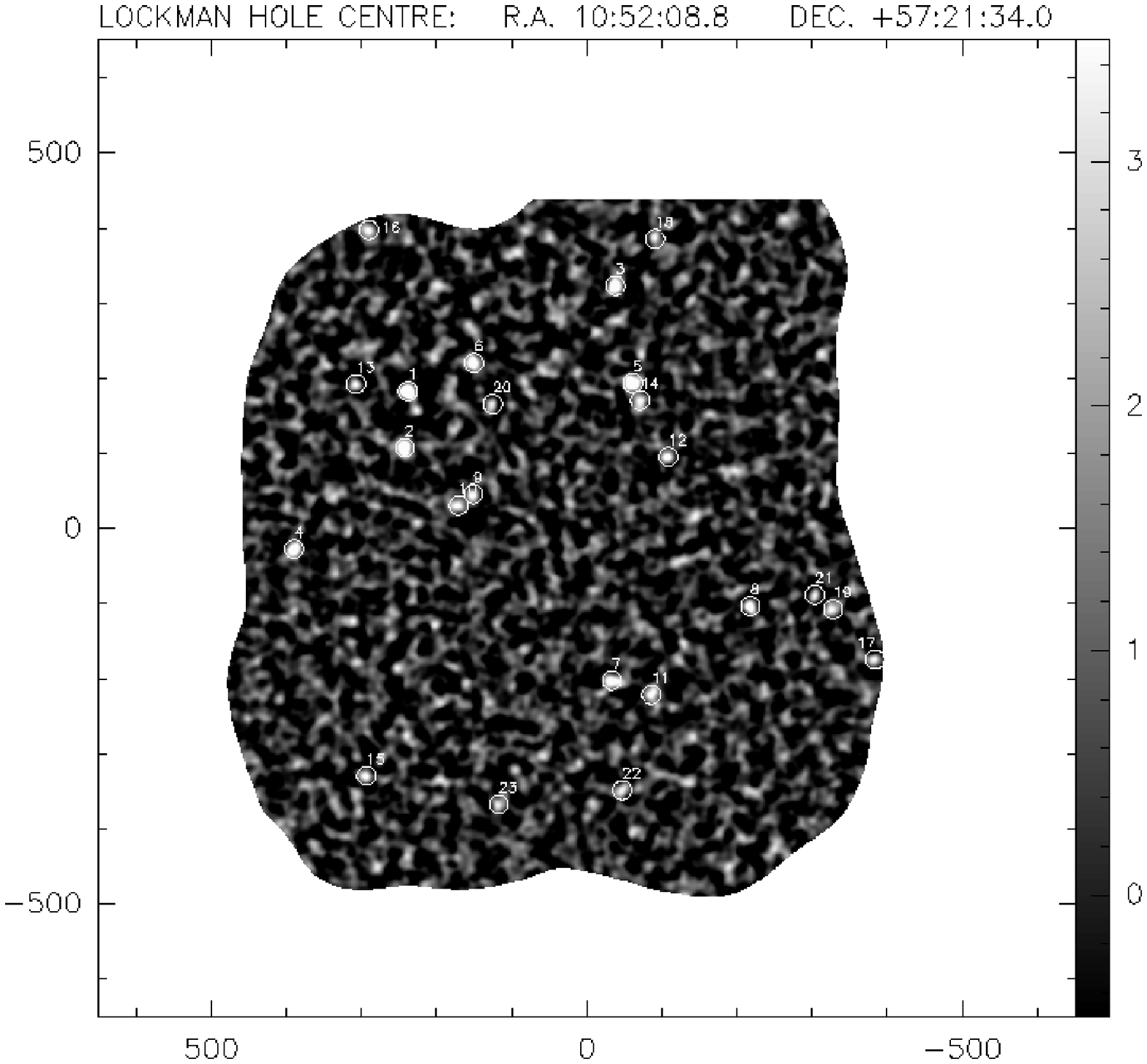}
\caption[$1200$\mum MAMBO signal-to-noise map of the Lockman Hole
field.]{$1200$\mum MAMBO signal-to-noise map of the Lockman Hole field. Sources
detected at $\ge$$3.5\sigma$ are circled in white, and are
numbered in order of significance. The axes denote the offset (arcsec) from
the map centre.}
\label{fig:map-lh}
\end{center}
\end{figure*}

Scan-mapping is the only option for large area mapping at
the IRAM 30m telescope, and as a result the chopping direction is not fixed on
the sky but varies with time which means that the chops are smeared out in the
final map. Hence, unlike SCUBA jiggle maps, one cannot utilise the extra
information contained in the position of the negative side-lobes for source
extraction. Instead, a simple matched-filtering technique was adopted, using a
Gaussian as the filter.  However, as is seen from the noise images shown in
Figure \ref{fig:noise-maps}, the noise is not entirely uniform across the two
fields and any attempt at extracting sources has to take this into account.
This was done by adopting a noise-weighted convolution technique similar to
that of Serjeant et al.~(2003).  In order to account for the 10.7$''$ beam and
the typical pointing error of 3$''$ rms, a Gaussian PSF, $P(x,y)$, with a {\sc
fwhm} of $\sqrt{10\parcsec7 ^2 + 3\parcsec0^2} = 11.1\arcsecs$ was fitted to
each pixel in the image $S(i,j)$ by minimising the following expression:
\begin{equation}
\chi^2 (i,j) = \sum_{x,y} W(i-x,j-y)\left ( S(i-x,j-y) - F P(x,y) \right ) ^2
\end{equation}
where $W$ is the weight image which is related to the noise by $W =
1/N^2$, and $F$ is the best-fit flux value to pixel $(i,j)$ and is
given by:
\begin{equation}
F(i,j) = \frac{\sum_{x,y} S(i-x,j-y)W(i-x,j-y)P(x,y)}{\sum_{x,y} W(i-x,j-y)P(x,y)^2}.
\end{equation}
The error on the flux can then be shown to be:
\begin{equation}
\Delta F(i,j) = \frac{1}{\sqrt{\sum_{x,y} W(i-x,y-j) P(x,y)^2}}.
\end{equation}
The signal-to-noise images, $F/\Delta F$, obtained in this way are shown in
Figure \ref{fig:map-n2} and \ref{fig:map-lh} for the ELAIS\,N2 field and
Lockman Hole, respectively.  The final source catalogues for ELAIS\,N2 and the
Lockman Hole are listed in Table~\ref{table:source-list-n2} and
\ref{table:source-list-lh}.  We find a total of 13 sources at significance $\ge$$4.0\sigma$, 
and 21 sources with $\ge$$3.5\sigma$ in the ELAIS\,N2 region while the
number of sources in the Lockman Hole are 14 and 23 at $\ge$$4.0\sigma$ and $\ge$$3.5\sigma$, respectively.

%
% Table 1 
%
\setcounter{table}{0}
\begin{table*}
%\scriptsize
\caption{$1200$\mum MAMBO source catalogue in the ELAIS\,N2 field. The
rms noise in the centre of the map is $\sim$0.65\,mJy\,beam\,$^{-1}$ and increases
towards the edges. The area surveyed in the ELAIS\,N2 field is 160
arcmin\,$^2$.}
\vspace{0.5cm}
\begin{center}
\begin{tabular}{llcccc}
\hline
\hline
ID          &   MAMBO ID    &   RA (J2000)  & Dec (J2000) &   $S_{1200\mu\rm{m}} \pm \sigma_{1200\mu\rm{m}}$   &   $S/N$ \\
            &               &             &                                  &         \\ \hline
\multicolumn{5}{c}{$\sigma \geq 4.0$ Detections} \\ \hline
MM\,J163647+4054 & N2\,1200.1   & 16:36:47.9  & +40:54:46  & $4.0 \pm 0.8$ & 5.00\\
MM\,J163639+4053 & N2\,1200.2   & 16:36:39.1  & +40:53:26  & $5.0 \pm 1.0$ & 5.00\\
MM\,J163635+4055 & N2\,1200.3   & 16:36:35.7  & +40:55:59  & $3.9 \pm 0.8$ & 4.87\\
MM\,J163639+4056 & N2\,1200.4   & 16:36:39.1  & +40:56:36  & $3.4 \pm 0.7$ & 4.85\\
MM\,J163632+4059 & N2\,1200.5   & 16:36:32.1  & +40:59:46  & $5.7 \pm 1.2$ & 4.75\\
MM\,J163708+4054 & N2\,1200.6   & 16:37:08.3  & +40:54:18  & $4.2 \pm 0.9$ & 4.66\\
MM\,J163640+4056 & N2\,1200.7   & 16:36:40.7  & +40:56:46  & $3.2 \pm 0.7$ & 4.57\\
MM\,J163710+4055 & N2\,1200.8   & 16:37:10.5  & +40:55:02  & $4.0 \pm 0.9$ & 4.44\\
MM\,J163705+4054 & N2\,1200.9   & 16:37:05.8  & +40:54:26  & $4.0 \pm 0.9$ & 4.44\\
MM\,J163650+4057 & N2\,1200.10  & 16:36:50.3  & +40:57:36  & $3.1 \pm 0.7$ & 4.42\\
MM\,J163656+4058 & N2\,1200.11  & 16:36:56.3  & +40:58:14  & $2.5 \pm 0.6$ & 4.16\\
MM\,J163713+4054 & N2\,1200.12  & 16:37:13.8  & +40:54:03  & $4.9 \pm 1.2$ & 4.08\\
MM\,J163619+4054 & N2\,1200.13  & 16:36:19.5  & +40:54:00  & $5.7 \pm 1.4$ & 4.07\\
\hline
\multicolumn{5}{c}{$4.0 > \sigma \geq 3.5$ Detections} \\ 
\hline
MM\,J163640+4058 & N2\,1200.14   & 16:36:40.4  & +40:58:44  & $3.1 \pm 0.8$ & 3.87\\
MM\,J163636+4057 & N2\,1200.15   & 16:36:36.2  & +40:57:19  & $3.1 \pm 0.8$ & 3.87\\
MM\,J163644+4102 & N2\,1200.16   & 16:36:44.8  & +41:02:01  & $2.7 \pm 0.7$ & 3.85\\
MM\,J163706+4053 & N2\,1200.17   & 16:37:06.7  & +40:53:15  & $4.2 \pm 1.1$ & 3.81\\
MM\,J163655+4059 & N2\,1200.18   & 16:36:55.9  & +40:59:12  & $2.2 \pm 0.6$ & 3.66\\
MM\,J163658+4104 & N2\,1200.19   & 16:36:58.3  & +41:04:37  & $2.9 \pm 0.8$ & 3.62\\
MM\,J163647+4055 & N2\,1200.20   & 16:36:47.9  & +40:55:39  & $2.5 \pm 0.7$ & 3.57\\
MM\,J163715+4055 & N2\,1200.21   & 16:37:15.6  & +40:55:40  & $3.9 \pm 1.1$ & 3.54\\
\hline
\label{table:source-list-n2}
\end{tabular}
\end{center}
\end{table*}

%
% Table 2  
%
\begin{table*}
%\scriptsize
\caption{$1200$\mum MAMBO source catalogue in the Lockman Hole field.  The rms
noise is $\sim$0.6\,mJy\,beam$^{-1}$ at the centre where the map is deepest,
and increases towards the edges. The total area surveyed is 197 arcmin$^2$.}
\vspace{0.5cm}
\begin{center}
\begin{tabular}{llcccc}
\hline
\hline
ID          &   MAMBO ID    &   RA (J2000)  & Dec (J2000) &   $S_{1200\mu\rm{m}} \pm \sigma_{1200\mu\rm{m}}$   &   $S/N$ \\
            &               &             &                                  &         \\ \hline
\multicolumn{5}{c}{$\sigma \geq 4.0$ Detections} \\ \hline
MM\,J105238+5724& LE\,1200.1   & 10:52:38.3  & +57:24:37  & $4.8 \pm 0.6$ & 8.00\\
MM\,J105238+5723& LE\,1200.2   & 10:52:38.8  & +57:23:22  & $4.1 \pm 0.6$ & 6.83\\
MM\,J105204+5726& LE\,1200.3   & 10:52:04.1  & +57:26:58  & $3.6 \pm 0.6$ & 6.00\\
MM\,J105257+5721& LE\,1200.4   & 10:52:57.0  & +57:21:07  & $5.7 \pm 1.0$ & 5.70\\
MM\,J105201+5724& LE\,1200.5   & 10:52:01.3  & +57:24:48  & $3.4 \pm 0.6$ & 5.66\\
MM\,J105227+5725& LE\,1200.6   & 10:52:27.5  & +57:25:15  & $2.8 \pm 0.5$ & 5.60\\
MM\,J105204+5718& LE\,1200.7   & 10:52:04.7  & +57:18:12  & $3.2 \pm 0.7$ & 4.57\\
MM\,J105142+5719& LE\,1200.8   & 10:51:42.0  & +57:19:51  & $4.1 \pm 0.9$ & 4.55\\
MM\,J105227+5722& LE\,1200.9   & 10:52:27.6  & +57:22:20  & $3.1 \pm 0.7$ & 4.42\\
MM\,J105229+5722& LE\,1200.10  & 10:52:29.9  & +57:22:05  & $2.9 \pm 0.7$ & 4.14\\
MM\,J105158+5717& LE\,1200.11  & 10:51:58.3  & +57:17:53  & $2.9 \pm 0.7$ & 4.14\\
MM\,J105155+5723& LE\,1200.12  & 10:51:55.5  & +57:23:10  & $3.3 \pm 0.8$ & 4.12\\
MM\,J105246+5724& LE\,1200.13  & 10:52:46.9  & +57:24:47  & $2.4 \pm 0.6$ & 4.00\\
MM\,J105200+5724& LE\,1200.14  & 10:52:00.0  & +57:24:25  & $2.4 \pm 0.6$ & 4.00\\
\hline
\multicolumn{5}{c}{$4.0 > \sigma \geq 3.5$ Detections} \\ 
\hline
MM\,J105245+5716& LE\,1200.15   & 10:52:45.1  & +57:16:05  & $3.1 \pm 0.8$ & 3.87\\
MM\,J105244+5728& LE\,1200.16   & 10:52:44.8  & +57:28:12  & $5.0 \pm 1.3$ & 3.84\\
MM\,J105121+5718& LE\,1200.17   & 10:51:21.5  & +57:18:40  & $4.8 \pm 1.3$ & 3.69\\
MM\,J105157+5728& LE\,1200.18   & 10:51:57.7  & +57:28:00  & $2.2 \pm 0.6$ & 3.66\\
MM\,J105128+5719& LE\,1200.19   & 10:51:28.4  & +57:19:47  & $4.0 \pm 1.1$ & 3.63\\
MM\,J105224+5724& LE\,1200.20   & 10:52:24.4  & +57:24:20  & $1.8 \pm 0.5$ & 3.60\\
MM\,J105131+5720& LE\,1200.21   & 10:51:31.3  & +57:20:06  & $3.6 \pm 1.0$ & 3.60\\
MM\,J105203+5715& LE\,1200.22   & 10:52:03.0  & +57:15:46  & $2.8 \pm 0.8$ & 3.50\\
MM\,J105223+5715& LE\,1200.23   & 10:52:23.4  & +57:15:27  & $2.8 \pm 0.8$ & 3.50\\
\hline
\label{table:source-list-lh}
\end{tabular}
\end{center}
\end{table*}

\subsection{Monte Carlo Simulations}\label{section:monte-carlos}

Extensive Monte Carlo simulations were performed in order to determine the
reliability of the source extraction technique, i.e.~how large a fraction of
the extracted sources are due to spurious noise peaks, and how well does the
extraction reproduce fluxes and positions as a function of signal-to-noise? Due
to the slight difference in survey depth and also geometry between the
ELAIS\,N2 and Lockman Hole maps, we decided to perform separate Monte Carlo
simulations for each of the two fields.

\begin{figure}
\begin{center}
\hspace*{-0.9cm}\includegraphics[width=1.1\hsize,angle=0]{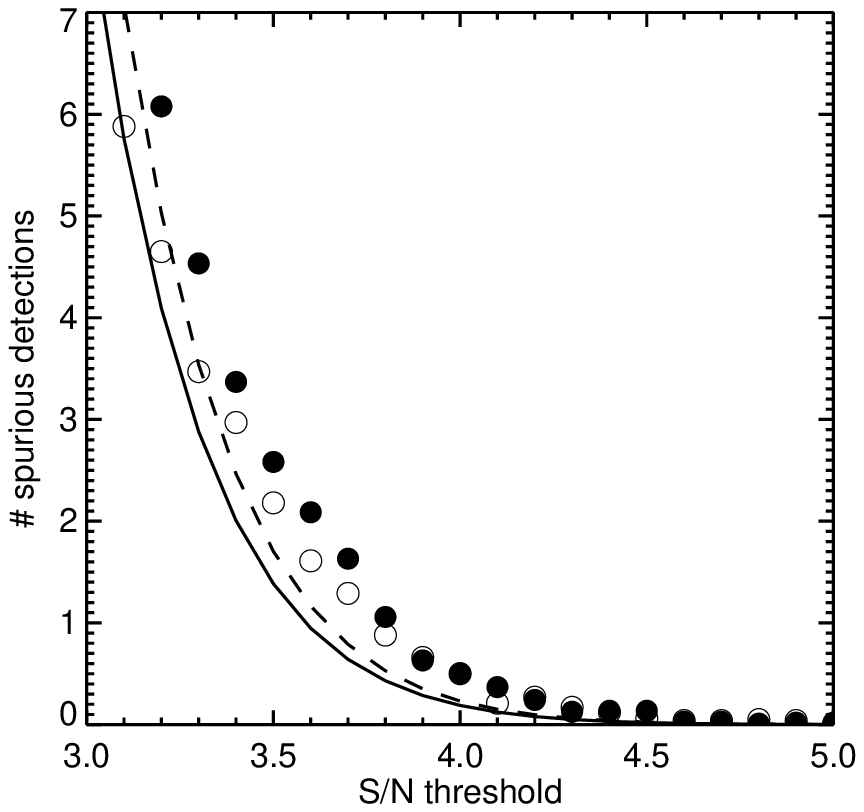}
\caption[Expected average number of spurious positive sources in the
ELAIS\,N2 $1200$\mum map as a function of signal-to-noise ratio.]{Expected
average number of spurious positive sources in ELAIS\,N2 (empty circles) and
the Lockman Hole (filled circles) as a function of signal-to-noise ratio.
The solid and dashed lines are the expected number of positive
spurious sources in the ELAIS\,N2 and Lockman Hole, assuming the noise is
purely Gaussian and the MAMBO beam is a Gaussian with a {\sc fwhm} of
$10.7\arcsecs$.  }
\label{fig:spurious-detections}
\end{center}
\end{figure}

In order to assess the contamination from spurious noise peaks to the source
catalogues, we produced maps based on the real data but with the astrometry
corrupted. This was done by randomising the array parameters of each scan and
then producing final maps using the same {\sc mopsi} data reduction pipeline as
for the real map. The advantage of this method is that the correlated noise in
the raw data is preserved, and is taken out during the data reduction process
in the same manner as for the real map. As a result, this method should give a
realistic picture of the number of spurious noise peaks expected in our maps.
We produced 100 such astrometrically corrupted maps and counted the number of
positive spurious sources as a function of detection threshold.  The results
are shown in Figure \ref{fig:spurious-detections} with empty and filled circles
referring to the ELAIS\,N2 and Lockman Hole maps, respectively.  As expected,
the number of spurious detections drops off exponentially as a function of
signal-to-noise threshold.  From Figure \ref{fig:spurious-detections} we find
that a threshold of $4.0\sigma$ results in less than one spurious source expected
at random, while a cut-off at $3.5\sigma$ yields less than 2 spurious sources
per field. We adopt $3.5\sigma$ as the detection cut-off for sources in both
fields since it provides a good compromise between catalogue size and source
reliability.

\begin{figure*}
\begin{center}
\includegraphics[width=0.8\hsize,angle=0]{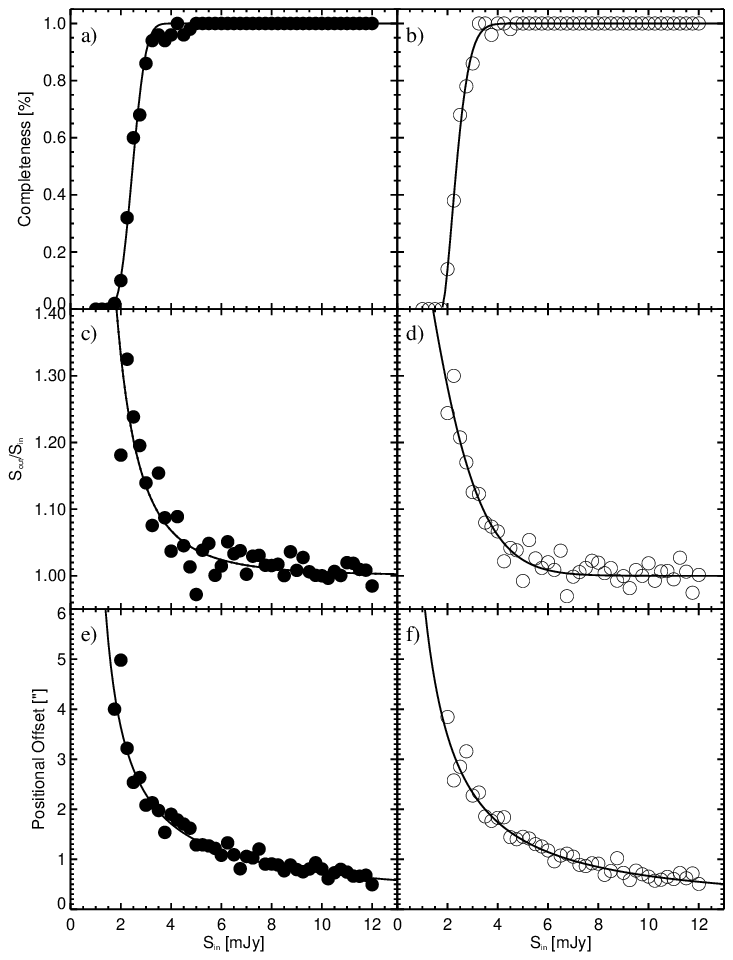}
\caption[Results of Monte Carlo simulations.]{Results from Monte Carlo
simulations. The solid lines in all three panels represent best fits to the
points, see text for details. As in Figure \ref{fig:spurious-detections}, empty
and filled circles refer to results from the ELAIS\,N2 and Lockman Hole maps.
\textbf{Top:} Percentage of added sources recovered against input flux in the
range 1 to 12\,mJy. \textbf{Middle:} The 'boosting' factor, i.e.\ the factor by
which the input flux is scattered upwards due to instrumental and confusion
noise (Eddington bias), plotted against input flux. \textbf{Bottom:} Deviations
between true (input) positions and the positions derived from the source
extraction technique as a function of input flux.}
\label{fig:completeness-flux-astrometry}
\end{center}
\end{figure*}

Monte Carlo simulations were also used to test the completeness and reliability
of the flux and position estimates. Sources in the flux range 1--12\,mJy were
added to the map in steps of 0.5\,mJy, one at a time and each at a random
position, with 50 sources in each flux bin.  We used a scaled beam pattern as
the template for the artificial sources.  However, due to sky-rotation, the
double-beam profile varies across the map and as a result a separate multi-beam
PSF had to be constructed for each source, depending on its position in the
map. To accomplish this, the source --- with its correct intensity and position
--- was added to the data stream and then the data were reduced in the same
manner as the real data. Thus, each artificial source added to the map had
exactly the right multi-beam profile corresponding to its position in the map.
Finally, sources were extracted using the same detection threshold as used for
the real map. If a simulated source happened to fall within half a beam width
of a real source in the map, it was discarded from the analysis. The results
from these Monte Carlo simulations are shown in Figure
\ref{fig:completeness-flux-astrometry}, again with open and filled circles
representing results based on the ELAIS\,N2 and Lockman Hole fields,
respectively.

Figures \ref{fig:completeness-flux-astrometry}a and b show the completeness of
the survey, i.e.~the percentage of recovered sources in our Monte Carlo
simulations as a function of input flux density. As expected the source
extraction does well in extracting all of the brighter sources, but fares less
well for sources close to the detection threshold.  In both fields the
completeness is seen to be $\sim$50 per cent at a flux level of
$\simeq$2.5\,mJy, increasing to about 95 per cent at 3\,mJy.  The solid curves
are best fits of the function $f(S_{in}) = 1 - exp(A(S_{in}-B)^C)$.

In Figures \ref{fig:completeness-flux-astrometry}c and d we have plotted the
ratio between the input and output fluxes $S_{out}/S_{in}$, as a
function of the input flux density. It is seen that fainter sources tend to
have higher extracted fluxes than brighter sources. This effect --- 'flux
boosting' in Scott et al.~(2002) --- is due to the instrumental noise from the
array itself and the confusion noise from faint sources below the detection
threshold conspiring to scatter the retrieved fluxes upwards. As expected, flux
boosting has the greatest impact at faint flux levels, where we find it can be
as high as $\sim$30 per cent. However, for input fluxes $\ge$2\,mJy the
boosting is on average no more than $\sim$20 per cent, which is comparable to
the calibration errors, see section \ref{sec:observations}.  It is also
comparable to the boosting factor reported by Eales et al.~(2003) from a
comparison of MAMBO fluxes obtained from a map and from photometry
observations. An average flux boosting factor of $\sim$15 per cent was derived
by Scott et al.~(2002) from Monte Carlo simulations of their 8\,mJy Survey.  The
solid lines in Figures \ref{fig:completeness-flux-astrometry}c and d represent
best fits to the data points of the function $f(S_{in}) = 1 + A
exp(BS_{in}^C)$.

The average positional offset between the input and output positions of the
added sources is shown as a function of flux density in Figures
\ref{fig:completeness-flux-astrometry}e and f. Not surprisingly, the source
extraction reproduces the true positions less well for faint sources than for
brighter sources. It is seen that, in the flux range 2--5\,mJy where most of
our sources lie, the positional error is of the order 1.5--3.0$''$. This is
slightly better than the 3--4$''$ errors typically quoted by SCUBA surveys
(Webb et al.~2003; Scott et al.~2002; Borys et al.~2003) and may be due to
the smaller MAMBO beam.

\section{Number counts of MAMBO sources}\label{section:number-counts-section}

In order to derive the $1200$\mum number counts we must correct the raw number
counts for the effects of flux boosting and the fact that the survey is not
complete at faint flux levels. Finally, we need to assess to what degree
spurious sources contaminate the number counts.

Addressing the last issue first, it is seen from Figure
\ref{fig:spurious-detections} that at a detection threshold of
$4.0\sigma$ we expect no spurious positive sources; at a threshold of $3.5\sigma$
we expect at most two. Furthermore, the flatness of the curves at
$3.5\sigma$ implies that a slight error in the noise estimate will not
change the number of false sources dramatically.  This would not have
been the case had we used a lower threshold of $3.0\sigma$. As a
result, we have used the $\ge$$3.5\sigma$ catalogues to derive the
number counts.  While we cannot predict which flux bins the spurious
sources will fall into, it is seen from Tables \ref{table:source-list-n2} and
\ref{table:source-list-lh} that the $\ge$$3.5\sigma$ sources span a wide range
in flux. It is therefore unlikely that the spurious sources will be restricted
to a single flux bin, and as a result their effect on the number counts
is expected to be negligible. 

Flux boosting was corrected for in a statistical way by applying the best-fit
curves in Figures \ref{fig:completeness-flux-astrometry}c and d to the raw
output fluxes. The boost-corrected number counts obtained in this way were then
corrected for completeness using the fitted completeness curves in Figures
\ref{fig:completeness-flux-astrometry}a and b.

\begin{figure*}
\begin{center}
\includegraphics[width=0.7\hsize,angle=0]{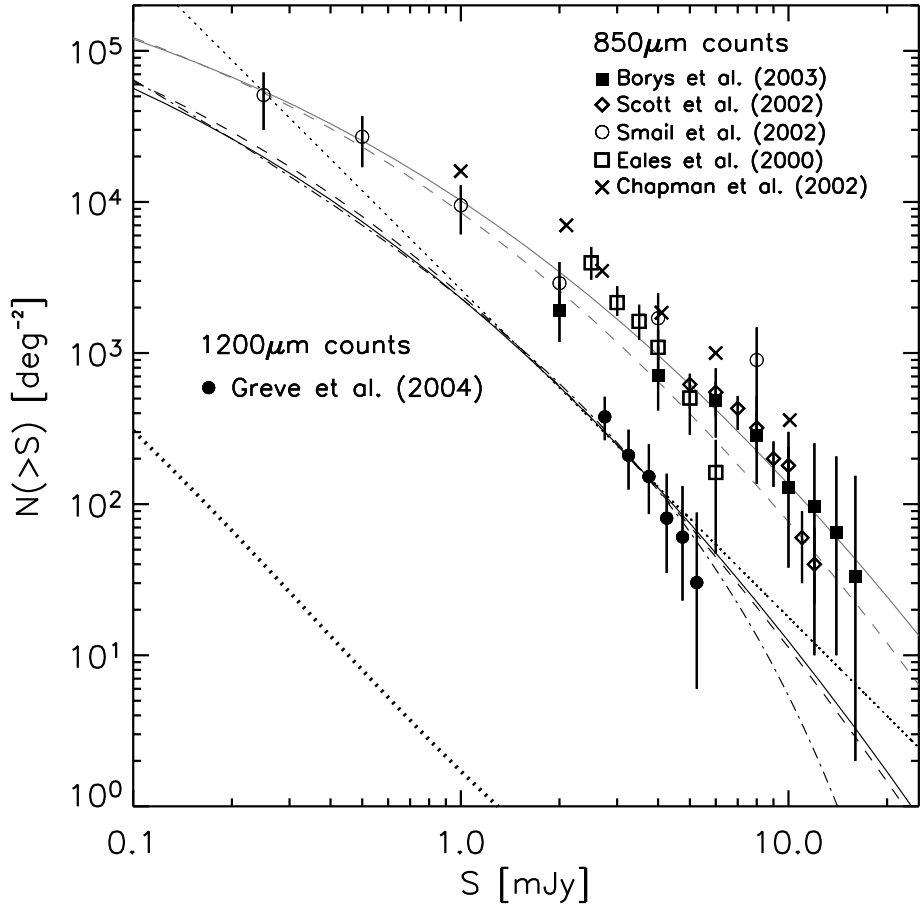} 
\caption[$1200$\mum cumulative number counts.]{Cumulative number counts
at $1200$\mum (solid circles) based on the $\ge$$3.5\sigma$ MAMBO source
catalogue presented in this paper. The error bars represent 95 per
cent two-sided confidence levels on a Poissonian distribution. Also
shown are $850$\mum cumulative number counts from a number of recent
SCUBA surveys. The dashed black and grey lines show the number counts at
$1200$ and $850$\mum, respectively, as predicted by a pure luminosity
scenario in which $g(z) = (1+z)^3$ out to $z\rm =2$, beyond which no
further evolution occurs.  The solid black and grey curves represent
predicted number counts at $1200$ and $850$\mumeol, respectively, based on
the galaxy evolution model by Jameson et al.~(1999), see text for
details. The thin dotted line corresponds to a simple power law with slope $\alpha = -2.2$, while
the dark solid curve represents the Schechter function $dN/dS \propto
(S/S_o)^{-\alpha}exp(-S/S_o)$, where $\alpha=2.3$ and $S_o = 4$\,mJy.
Finally, the thick dotted line represent a no-evolution scenario.}
\label{fig:number-counts-figure}
\end{center}
\end{figure*}

The final integrated number counts in flux bins 2.75, 3.25, 3.75, 4.25,
4.75, and 5.25\,mJy are given in Table \ref{table:number-counts-table} along with the
raw number counts. The table shows the counts derived for the ELAIS\,N2 and
Lockman Hole separately, as well as the combined number counts. The quoted
errors correspond to the 95 per cent two-sided confidence level of a Poissonian
distribution.  The number counts derived from the two fields separately are
seen to agree well within the error. In Figure \ref{fig:number-counts-figure} we have
plotted the corrected accumulative $1200$\mum number counts as derived from the
MAMBO survey presented in this paper.  While the surface density of SCUBA
sources detected at $850$\mum has been fairly well constrained over a large range
in flux density thanks to a number of large submm surveys (Smail et al.~1997; Scott et al.~2002), 
only one other published MAMBO survey has so far attempted
to constrain the $1200$\mum number counts (Bertoldi et al.~2000). 

For comparison we have also plotted the $850$\mum source counts as derived from
a number of SCUBA surveys. It is seen that the $1200$\mum counts are lower than
those at $850$\mumeol. This is expected if one assumes that the MAMBO and SCUBA
sources trace the same population of dust-enshrouded galaxies: the
lower MAMBO counts are due to the fact that one is sampling further down the
Rayleigh-Jeans tail than at $850$\mumeol. Thus, if one scales the MAMBO fluxes
by a factor of $\sim$2.5, which roughly corresponds to the $850/1200$\mum flux
ratio for a starburst at $z\sim\rm 2.5$, the MAMBO counts are found to coincide
with the SCUBA counts.  The overall similarity between the shape of the $1200$
and $850$\mum number counts lends support to the view that the MAMBO and SCUBA
sources are the same population viewed at slightly different wavelengths.

At flux levels fainter than $\sim$4\,mJy the MAMBO counts display a power-law
slope of $\alpha \simeq -2.2$, similar to that of the SCUBA counts in the flux
range 2--9\,mJy.  The $1200$\mum source counts appears to show a break at
$\sim$4\,mJy beyond which the slope of the counts steepens to $\alpha \sim
-4.7$. The dot-dashed curve in Figure
\ref{fig:number-counts-figure} represents a Schechter-type function of
the form $dN/dS \propto (S/4\rm{mJy})^{-2.3}exp(-S/4\rm{mJy})$. The
Schechter function has a natural break at 4\,mJy and appears to provide a 
somewhat better match than a power law, although the data do not allow for
a statistically significant distinction between a power law and a Schechter function.
At the bright end of the $850$\mum counts there is some
disagreement between the SCUBA surveys, with the Scott et al.~(2002) counts
dropping off more steeply than found by Borys et al.~(2003). The $1200$\mum
counts show tentative evidence for a break at a flux level which corresponds to
the $850$\mum flux at which the SCUBA counts by Scott et al.~(2002) appear to
turn over, suggesting that this feature is real. In order to better constrain
the bright end of the number counts, larger surveys such as the one square degree
MAMBO Deep Field Survey (Bertoldi et al., in preparation) and SHADES are
needed. Confirmation of our tentative findings would be important --- the
bright (and very faint) ends of the number counts hold the biggest potential in
terms of discriminating between models.

At $1200$\mum the extragalactic background amounts to $I_{\nu}\nu \sim 0.16$\,nW\,m$^2$\,sr$^{-1}$
(Fixsen et al,~1998). By integrating up $S dN/dS$ over the flux range 2.25--5.75\,mJy
using the above Schechter function, we estimate that our survey has resolved about
10 per cent of the background light at $1200$\mumeol. This is comparable to the UK 8\,mJy
Survey which also resolved $\sim 10$ per cent of the extragalactic background at $850$\mumeol. 

%
% Table 3 
%
\begin{table*}
%\scriptsize
\caption{$1200$\mum source counts as derived from the ELAIS\,N2 field
(columns 2--4), Lockman Hole (columns 5--7), and the combined fields
(columns 8--10). Only the $\ge$$3.5\sigma$ catalogue is used.  The area
surveyed in the three cases are 160, 197, and 357 arcmin$^2$,
respectively.  The errors corresponds to 95 per cent two-sided confidence
levels of a Poissonian distribution.}
\vspace{0.5cm}
\begin{center}
\begin{tabular}{cccccccccccccc}
\hline
\hline
          &                & {\bf ELAIS\,N2}  &                & & &               & {\bf LOCKMAN}   &                & & &               & {\bf TOTAL}     &               \\
$S$       &  $N_{raw}(>S)$ & $N(>S)$    & $N(>S)$        & & & $N_{raw}(>S)$ & $N(>S)$   & $N(>S)$        & & & $N_{raw}(>S)$ & $N(>S)$   & $N(>S)$       \\
          &                & corrected  & corrected      & & &               & corrected & corrected      & & &               & corrected & corrected     \\
$[$mJy$]$ &                &            & $[$deg$^{-2}]$ & & &               &           & $[$deg$^{-2}]$ & & &               &           & $[$deg$^{-2}]$\\
\hline
2.75      & 16             & 19.99      & $450^{+218}_{-187}$   & & & 14            & 17.49     & $320^{+178}_{-136}$     & & &  30           & 37.48     &$378^{+136}_{-113}$\\
3.25      & 11             & 11.47      & $258^{+185}_{-134}$   & & & 9             & 9.39      & $172^{+141}_{-96}$      & & &  20           & 20.86     &$210^{+101}_{-85}$\\
3.75      &  9             & 9.05       & $204^{+181}_{-111}$   & & & 6             & 6.03      & $110^{+128}_{-70}$      & & &  15           & 15.08     &$152^{+98}_{-66}$\\
4.25      &  4             & 4.00       & $90^{+140}_{-65}$     & & & 4             & 4.00      & $73^{+114}_{-53}$       & & &  8            & 8.00      &$81^{+78}_{-46}$\\
4.75      &  4             & 4.00       & $90^{+140}_{-65}$     & & & 2             & 2.00      & $37^{+96}_{-32}$        & & &  6            & 6.00      &$61^{+71}_{-38}$\\
5.25      &  2             & 2.00       & $45^{+117}_{-40}$     & & & 1             & 1.00      & $18^{+84}_{-18}$        & & &  3            & 3.00      &$30^{+58}_{-24}$\\
\hline
\label{table:number-counts-table}
\end{tabular}
\end{center}
\end{table*}

If indeed MAMBO and SCUBA sources are the same, then it should be
possible to construct models which can simultaneously reproduce the
$850$ and $1200$\mum counts.

In order to do so, we considered a simple parametric model which is
based on the local $60$\mum LF, $\Phi_o(L_{60})$, as derived from
{\em IRAS} data (Saunders et al.~1990).  The latter provides the best
estimate of the far-IR LF of dusty galaxies in the local Universe to
date, and is thus a useful template to try and model the number counts
of dusty high-redshift galaxies.  While a local $850$\mum LF has been
established by Dunne et al.~(2000), who used SCUBA to observe a large
number of galaxies from the {\em IRAS} Bright Galaxy Sample, the
bright end of this function is still highly uncertain. Furthermore, if
the majority of the MAMBO sources are at redshifts $z\rm >2$ and have
similar dust temperatures to the {\em IRAS} galaxies (30--40\,{\sc
k}), then the $60$\mum LF will be a better approximation to the
rest-frame far-IR LF of the MAMBO sources.  We have therefore adopted
the $60$\mum LF as the reference LF at $z\rm =0$.  In order to assess
the evolution in the far-IR LF, we used a simple approach in which the
LF at redshift $z$ is given by a simple scaling and/or translation of
the local LF, i.e.~$\Phi(L,z) = f(z)\Phi_o(L/g(z))$, where $f(z)$ and
$g(z)$ are parametric models of the evolution in number density and
luminosity, respectively. It is then straightforward to compute the
cumulative source counts per unit solid angle brighter than a given
flux density, $S_{\nu}$, as
\begin{equation}
N(\geq S_{\nu}) = \int_0^{z_{max}}\int_{L_{min}(S_{\nu},z)}^\infty
                  f(z) \Phi_o \left ( \frac{L}{g(z)} \right ) dlogL \frac{dV_c}{dz}dz,
\end{equation}
where $dV_C/dz$ is the co-moving volume element per redshift
increment, and $L_{min}(S_{\nu},z) = 4 \pi D^2_L S_{\nu} (1+z)^{-1} L_{\nu}/L_{\nu (1+z)}$
is the minimum luminosity observable for a
source at redshift $z$ and a survey flux limit of $S_{\nu}$.  The
evolution functions $f(z)$ and $g(z)$ are constrained by the fact that
their predictions of the extragalactic background, the observed number
counts and redshift distribution has to be consistent with
observations.

In Figure \ref{fig:number-counts-figure} we have plotted the predicted $1200$
and $850$\mum number counts from a model with a luminosity evolution of
$g(z) = (1+z)^3$ out to $z\rm =2$, beyond which the evolution is
unchanged out to $z\rm =10$, which marks the high-redshift cut-off of
the model. The model assumes the same SED for all sources --- here we
have adopted $T_{\rm d} = 44$\,{\sc k}, $\beta = +1.2$, and a critical
frequency of $\nu_c = 2$\,THz. While this pure-luminosity evolution
model does an excellent job at reproducing the number counts at
$1200$\mum it fares less well at $850$\mumeol, where it tends to slightly
underpredict the number counts at bright flux levels.

Another parametric model is that of Jameson (1999) which employs a
pure luminosity evolution of the form
\begin{equation}
g(z) = (1+z)^{3/2} sech^2 \left ( b \ln (1+z) - c \right ) cosh^2 c,
\label{eq:jamenson-g}
\end{equation}
where $b = 2.2\pm 0.1$ and $c=1.84\pm 0.1$, see also Smail et al.\
(2002). This model is arguably the most realistic of the parametric
models, since it is motivated by semi-analytical models, i.e.~models
based on dark matter halo merging trees and assumptions about the
astrophysics of the gas in halos.  Furthermore, the model is not only
in agreement with predictions of the chemical enrichment as a function
of cosmic time but it also naturally includes a peak in the evolution
at $z\rm \simeq 2$ which is in agreement with the recently determined
redshift distribution of radio-identified SMGs (Chapman et al.\
2003). In Figure \ref{fig:number-counts-figure} we have plotted the predicted
$1200$ and $850$\mum number counts for this model, under the assumption
that the SEDs of all MAMBO and SCUBA sources are well matched by
modified blackbody law with $T_{\rm d} = 37$\,{\sc k}, $\beta=+1.5$,
and a critical frequency of $\nu_c = 2$\,THz. It is seen that this
physically more realistic model is able to reproduce both the MAMBO
and SCUBA number counts extremely well, suggesting that the MAMBO and
SCUBA sources trace the same population of high-redshift,
far-IR-luminous, starburst galaxies.

While the mm number counts at faint as well as bright flux levels is
still too poorly determined to warrant a detailed test of models of
galaxy evolution, it is clear from the comparison with simple
analytical model made above, that a scenario in which no evolution
takes place at all, as illustrated by the thick dotted line in Figure
\ref{fig:number-counts-figure}, can be ruled out.

\section{Clustering of MAMBO sources}\label{section:clustering}

The next big step forward in our understanding of SMGs is likely to
come from determining how they are clustered. Their clustering
properties may then provide a link to a present-day population of
galaxies.  If SMGs are the progenitors of massive elliptical galaxies,
as suggested by their star-formation rates, molecular/dynamical masses
and co-moving space densities, then they are expected to be strongly
clustered. This follows from the way peaks in the density field in the
early Universe are biased in mass (e.g.~Benson et al.~2001).  Giant
ellipticals, being the most luminous objects in the local Universe,
are often found residing in the centres of galaxy clusters; as such
they pin-point the most overdense and therefore mass-biased regions in
the Universe.

Submm surveys of various depths and sizes have searched for clustering
among SMGs and have all failed to detect a significant signal (Scott
et al.~2002; Webb et al.~2003; Borys et al.~2003).  This has been
due largely to the limited size of the survey regions, though efforts
are further hampered by the fact the (sub)mm population spans a broad
range in redshift --- the quartile range is $z$ = 1.9--2.8 (Chapman et
al.~2003, 2004) with a possible high-redshift tail extending to $z\rm
\gs 4$ --- and any clustering signal will thus, when projected onto
the sky, become heavily diluted. Any detection of angular clustering
will therefore be a lower limit on the real 3D clustering. One lesson
learned from these surveys was that very large ($\sim$1 square degree)
areas containing several hundred sources with strong redshift
constraints are required in order to determine the clustering
properties of SMGs. Here, we present the results for the 
the MAMBO population using two independent clustering statistics.

The first test calculates the angular two-point correlation function,
$w (\theta)$, which quantifies the excess probability of finding
a source within an angle, $\theta$, of a randomly selected source,
over that of a random distribution, i.e.~
\begin{equation}
\delta P(\theta) = N^2 ( 1 + w(\theta) ) \delta \Omega_1 \delta \Omega_2,
\end{equation}
where $\delta P(\theta)$ is the probability of finding a source within
a solid angle $\delta \Omega_1$, and another source in another solid
angle $\delta \Omega_2$ within a angular distance $\theta$ of each
other (see e.g.~Coles \& Lucchin~2002).  $N$ is the mean surface
density of objects on the sky.

Several estimators of $w (\theta)$ have been suggested in the
literature, and here we shall use the one first proposed by Landy \&
Szalay (1993):
\begin{equation}
w(\theta) = \frac{\langle DD\rangle - 2\langle DR\rangle + \langle
RR\rangle}{\langle RR\rangle},
\end{equation}
where $\langle DD\rangle$ is the number of real source pairs which
fall within a bin of width $\delta \theta$ in the map.  $\langle RR
\rangle$ is the number of random--random pairs extracted from
simulated maps in which the sources are randomly distributed.
Similarly, $\langle DR\rangle$ is the number of data--random pairs.
In order to generate the random catalogues we created simulated maps
by drawing sources from a source count model and placing them randomly
throughout the field.  We used the best fit to the observed number
counts as provided by the Schechter function in section
\ref{section:number-counts-section}. Noise was added using one of the pure
noise maps obtained by randomising the array parameters (section
\ref{section:monte-carlos}) thereby ensuring that the noise mimicked
the properties of the real map as closely as possible. Mock source
catalogues were then generated by applying our source extraction
technique to the simulated maps. This exercise was repeated 500 times.
By taking the ensemble average, the catalogues $\langle RR\rangle$ and
$\langle DR\rangle$ were obtained. Furthermore, the $\langle
RR\rangle$ and $\langle DR\rangle$ catalogues were normalised to have
the same number of objects as $\langle DD\rangle$.  The clustering
analysis described here is similar to that of Borys et al.~(2003) in
the sense that our analysis takes the negative off-beams properly into
account when estimating the clustering: our simulated maps have the
same global chop-pattern as the real map. However, this effect is
expected to be very small for MAMBO maps, where the off beams are
smeared out due to sky rotation.

\begin{figure}
\begin{center}
\includegraphics[width=1.0\hsize,angle=0]{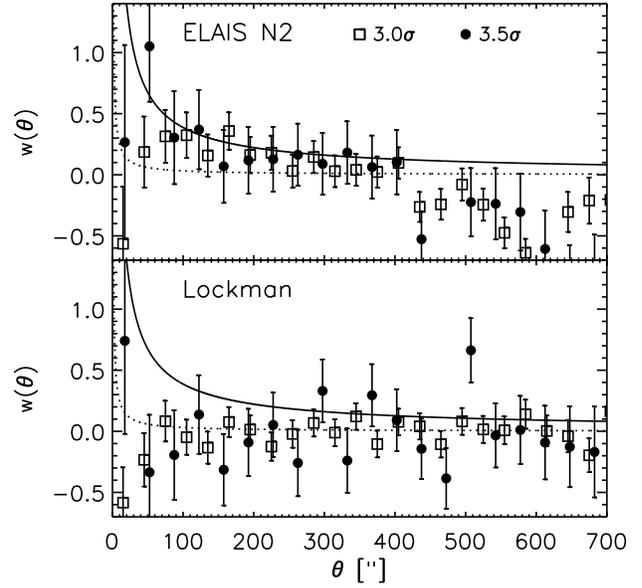} 
\caption[Angular two-point correlation function for ELAIS\,N2 (top)
and the Lockman Hole (bottom).]  {Angular two-point correlation
function for ELAIS\,N2 (top) and the Lockman Hole (bottom). The
two-point correlation function based on the $\ge$$3.0\sigma$ sample
(empty squares) uses a bin size of $30\arcsecs$, while the correlation
function based on the $\ge$$3.5\sigma$ catalogue (filled circles) is
derived using a bin size of $35\arcsecs$. The solid line is the
best-fit power-law correlation function found by Daddi et al.~(2000)
for EROs with $R-K >\rm 5$ and $K <\rm 18.5$. The dotted line
represents the correlation function for LBGs as determined by
Giavalisco \& Dickinson (2001).  The errors are given by $\delta
w(\theta) = \left ( \frac{1 + w (\theta)}{\langle DD\rangle}
\right )^{1/2}$.}
\label{fig:2pc}
\end{center}
\end{figure}

The resulting two-point correlation functions obtained using the
$\ge$$3.5\sigma$ as well as the $\ge$$3.0\sigma$ catalogues for the
ELAIS\,N2 and Lockman Hole fields are shown in Figure
\ref{fig:2pc}. The bin sizes used were $\delta \theta = 35\arcsecs$
and $30\arcsecs$, respectively, both of which roughly corresponds to
two times the MAMBO beam and furthermore ensures an adequate number of
sources in each bin. It is seen from Figure \ref{fig:2pc} that the
two-point correlation function for the Lockman Hole is consistent with
zero within the error bars at all angular scales.  In the ELAIS\,N2
field, however, the correlation function has a gradient, showing a
slight excess correlation at small angular scales and an
anti-correlation at scales of $\sim 500\arcsecs$. The correlation
functions based on the $\ge$$3.5\sigma$ and $\ge$$3.0\sigma$
catalogues both show this trend. The correlation seen at $\theta \ls
100\arcsecs$ seems to be apparent by looking at the map, from which it
appears that the sources are distributed in 'clusters', one of which
is located $\sim$200$\arcsecs$ south-west of the map centre, and
another $\sim$350$\arcsecs$ south-east from centre.  Within these
clusters, the sources are typically separated by $\sim$20--80$\arcsecs$, 
which could explain the excess correlation seen on
these scales.  The anti-correlation at $\theta \sim 500\arcsecs$ is
reflected in the overall distribution of sources in the ELAIS\,N2
field which shows the sources to be distributed around a void slightly
north-east of the map centre.  A similar void, although less
significant, is seen south-east of the centre of the Lockman Hole
map. These voids appear to be real: not only are there no sources
detected in these regions with SCUBA, but the $\mu$Jy radio population
also seems to have a similar spatial distribution (Ivison et al.~2002).
The good correspondence between these three populations
suggests that the structure seen is real large-scale structure. A more
detailed analysis of this will be presented in a future paper (Greve
et al., in preparation).

\begin{figure}
\begin{center}
\includegraphics[width=1.02\hsize,angle=0]{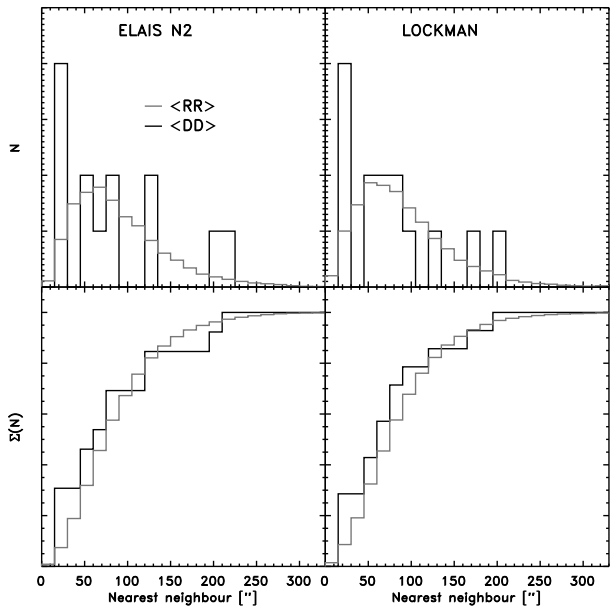} 
\caption[Nearest-neighbour clustering analysis of MAMBO sources in the
ELAIS\,N2 and Lockman Hole.]  {Distributions of nearest-neighbours in the
ELAIS\,N2 (left column) and the Lockman Hole (right column).
The black curves represent the actual data (the $\ge$$4.0\sigma$
source catalogues), while the red curves are the ensemble averaged distributions
obtained from 500 simulated maps of the two fields.  }
\label{fig:nn}
\end{center}
\end{figure}

An alternative test for clustering is the so-called nearest-neighbour
analysis (Scott \& Tout 1989). This considers the distribution of
nearest neighbour separations, $\theta$, of sources on a sphere. For a
random distribution of sources, the probability density function can
be shown to be:
\begin{equation}
P(\theta)d\theta = \frac{N-1}{2^{N-1}}\sin \theta (1+\cos \theta)^{N-2} d\theta,
\label{eq:nn}
\end{equation}
where $N$ is the number of objects in the sample --- see Scott \& Tout
(1989) for details. We performed a nearest-neighbour analysis on the
$\ge$$4.0\sigma$ source catalogues in each field, and the resulting
nearest-neighbour distributions are shown as black lines in Figure
\ref{fig:nn}. First of all, it is notable that the nearest-neighbour
distributions for the two fields look remarkably similar. In
particular, both distributions show a significant peak at angular
scales of $\sim$23$\arcsecs$.  Changing the bin size to values in the
range 11--17$\arcsecs$ does not alter the overall appearance of the
distributions significantly, and the over-density of sources in the
15--30$\arcsecs$ bin remains.  In order to compare the angular
distribution of MAMBO sources with a random distribution, we used the
same 500 $\langle RR\rangle$-catalogues which were used in connection
with the two-point correlation function.  The ensemble-averaged
nearest-neighbour distributions for these simulated maps are shown as
grey curves in Figure \ref{fig:nn}.

It is seen that the random distributions do not peak at
$\sim$15--30$\arcsecs$ but at $\theta \simeq 60\arcsecs$ which is in
accordance with the probability density function in eq.~\ref{eq:nn}.
From a Kolmogorov-Smirnoff (K--S) test we find that in both fields 
the probability of the $\langle DD\rangle$ and $\langle RR\rangle$ 
distributions being drawn from the same underlying distribution 
is about 10 per cent. In other words, there is a 10 per cent
chance of a random distribution yielding a more paired
distribution than that observed for the MAMBO sources. 
A similar analysis on the $\ge$$3.5\sigma$ sample shows a
similar significant peak at $\sim$15--30$\arcsecs$. 

Thus, while we haven't found a significant clustering signal from the two-point
correlation function, there is tentative evidence from the nearest-neighbour
analysis that MAMBO sources are not randomly distributed but tend to come in
pairs. This is qualitatively in agreement with a recent study which utilises
the results from the spectroscopic survey of radio-bright SMGs to search for
pairs and/or triplets of sources in redshift space. This has yielded a
significant detection of the clustering of SMGs and has constrained the
correlation length to $r_0 \simeq 6.1\pm 2.1$h$^{-1}$\,Mpc (Blain et al.~2004; Smail
et al.~2003).

\section{Comparison with the SCUBA UK 8\,mJy Survey}\label{section:mambo-scuba}

While a detailed analysis of the properties of the MAMBO sources at
radio and optical/near-IR wavelengths shall be presented in Paper II
(Greve et al., in preparation), it is appropriate here to compare the
MAMBO and SCUBA maps. Such a comparison is meaningful since as we saw
in section \ref{section:number-counts-section} both surveys reach very similar integral
counts, and a 'typical' $z\sim 2.5$ starburst SED will hit the flux
limit in both surveys almost simultaneously.

\subsection{The reliability of (sub)mm surveys}\label{subsection:comparison}

First of all, such a cross-check between the MAMBO and SCUBA
catalogues will allow us to assess the reliability of (sub)mm surveys.
As already mentioned, (sub)mm surveys in general have adopted
detection thresholds at low signal-to-noise ratios (typically
$3.0$--$3.5\sigma$ - see e.g.~Eales et al.~2000; Scott et al.~2002;
Webb et al.~2003; this work), and confusion has been a major issue
for these surveys. In addition, different surveys have used different
data-reduction software, and widely differing source extraction
techniques have been adopted.

\begin{figure*}
\begin{center}
\includegraphics[width=0.9\hsize,angle=0]{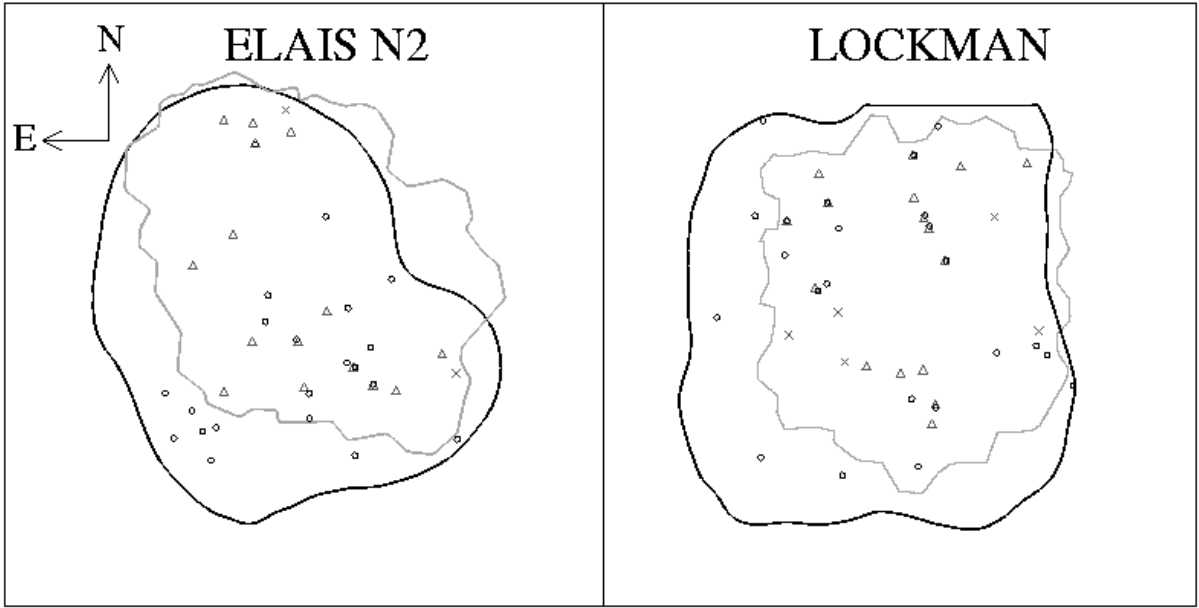} 
\caption[Distribution of MAMBO and SCUBA sources detected at
$\ge$$3.5\sigma$ in the ELAIS\,N2 and Lockman Hole fields (this paper;
Scott et al.~2002).]  {Distribution of MAMBO and SCUBA sources
detected at $\ge$$3.5\sigma$ in the ELAIS\,N2 and Lockman Hole fields
(this paper; Scott et al.~2002).  The black circles are sources
detected by MAMBO while grey triangles are SCUBA detections. The
crosses denote the seven SCUBA sources which were deemed spurious and
therefore rejected by Ivison et al.~(2002) due to their lack of radio
identifications and high associated noise levels.  The MAMBO and SCUBA
survey regions are outlined in black and grey, respectively, see also
Figure \ref{fig:mambo_scuba_overlay}. The boxes are $21\parcmin7\times 21\parcmin7$.}
\label{fig:mambo_scuba_overlay_sources}
\end{center}
\end{figure*}

The comparison between the MAMBO and SCUBA $\ge$$3.5\sigma$ source
catalogues is shown in Figure \ref{fig:mambo_scuba_overlay_sources},
which shows the outline of the MAMBO and SCUBA maps with the
$\ge$$3.5\sigma$ $850$\mum sample as given by Scott et al.~(2002) and
the $\ge$$3.5\sigma$ $1200$\mum sources presented in this paper
overplotted. It is seen immediately that four SCUBA sources in
ELAIS\,N2 are unambiguously detected at $1200$\mumeol, while in the
Lockman Hole eight SCUBA sources have been confirmed with MAMBO.

\begin{figure}
\begin{center}
\includegraphics[width=1.0\hsize,angle=0]{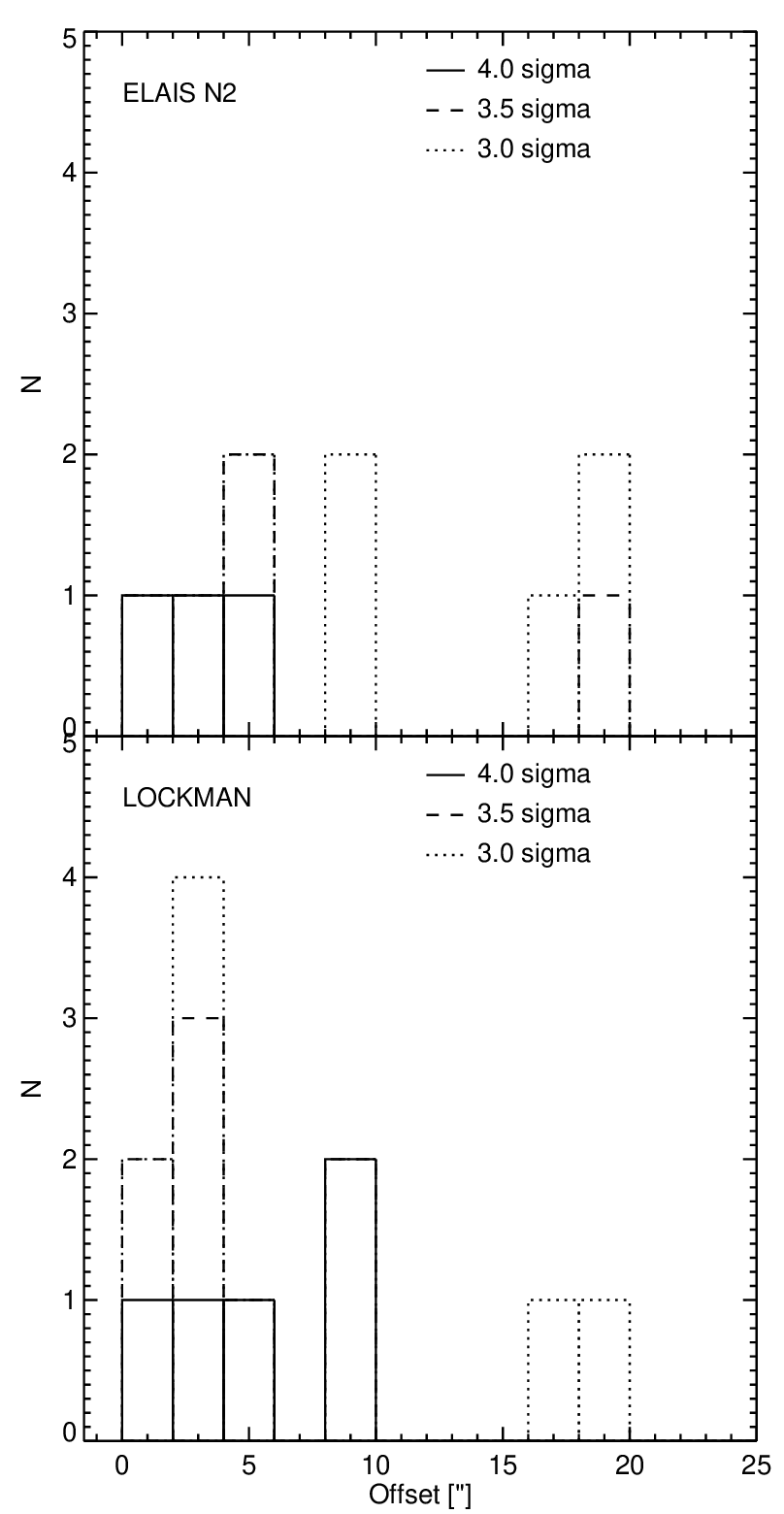}
\caption[Distribution of offsets between MAMBO and SCUBA sources.]
{Distribution of offsets between SCUBA sources and their nearest MAMBO
counterpart in the ELAIS\,N2 (top panel) and Lockman Hole fields
(bottom panel). The $4.0$, $3.5$, and $3.0\sigma$ MAMBO samples are
compared with the $4.0$, $3.5$, and $3.0\sigma$ SCUBA samples of Scott et
al.~(2002), and the corresponding distributions are shown as solid,
dashed and dotted curves, respectively.  }
\label{fig:scubo-dist}
\end{center}
\end{figure}

It is interesting to see how these MAMBO identifications are
distributed in terms of signal-to-noise and whether the identification
rate increases if we lower the source detection threshold to
$3.0\sigma$.  In order to do so, we have compared the $4.0$, $3.5$, and
$3.0\sigma$ SCUBA source lists of Scott et al.~with our corresponding
MAMBO source catalogues, and for each SCUBA source we computed the
distance to the nearest MAMBO source.  The resulting distributions are
shown in Figure \ref{fig:scubo-dist}.  For both fields the
distribution is seen to peak at offsets smaller than 10$\arcsecs$. In
fact, both distributions seem to show that a cutoff in positional
offset of $< 10\arcsecs$ is a natural selection criterion as to
whether a MAMBO source is a genuine counterpart to a SCUBA source or
not. This is in line with what we would expect given the {\sc fwhm}s
of the MAMBO and SCUBA beams.  Adopting this criterion and using the
$\ge$$3.0/3.5/4.0\sigma$ catalogues from both surveys, we confirm 6/4/3
out of 36/17/7 SCUBA sources in the ELAIS\,N2 field and 9/8/5 out of
36/21/12 SCUBA sources in the Lockman Hole, respectively. Thus, the
identification rate clearly increases with signal-to-noise, from
17--25 per cent for the $3.0\sigma$-catalogues to 43 per cent in
ELAIS\,N2 and 42 per cent in the Lockman Hole for the
$4.0\sigma$-catalogues.  At face value, our MAMBO survey thus confirms
about half of the most significant SCUBA sources.

\begin{figure}
\begin{center}
\includegraphics[width=0.95\hsize,angle=0]{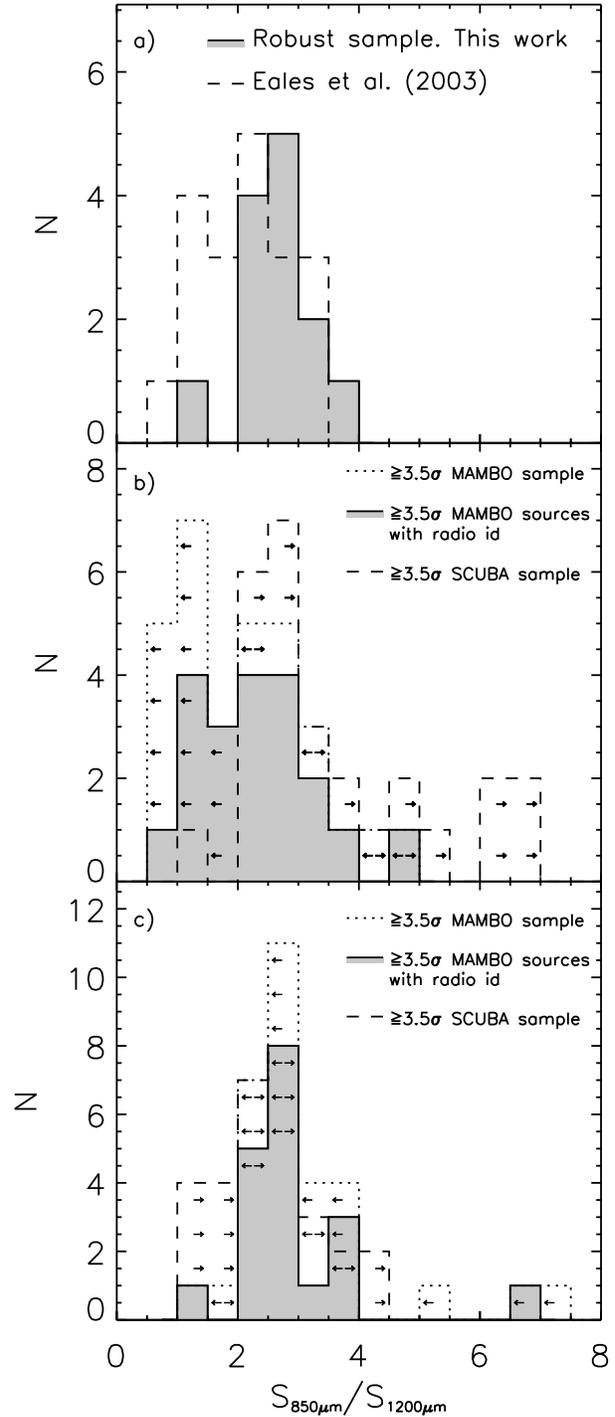}
\caption[$850/1200$\mum flux density ratio for SCUBA sources detected by
MAMBO.]  { {\bf a)} The distribution of $850/1200$\mum flux ratios for the 13
$\ge$$3.5\sigma$ SCUBA sources identified with MAMBO in this paper (light-grey
shaded area), and for the sample presented by Eales et al.~(2003) (dashed
line). {\bf b)} The dotted histogram is the flux ratio distribution for all the $\ge 3.5\sigma$ MAMBO
sources within the SCUBA regions, including the ones which were not detected
by SCUBA. In the latter case upper flux limits were estimated as the peak flux
within an aperture of radius 10\arcsecs centered on the MAMBO position in 
the SCUBA map. The distribution of flux ratios 
for all the $\ge 3.5\sigma$ SCUBA sources (dashed histogram) were obtained
in a similar manner by measuring the peak flux in the MAMBO map within a 10\arcsec
radius of the SCUBA position. The light-grey shaded histogram represents the
distribution for the subset of MAMBO sources which were robustly identified in 
the radio.
{\bf c)} The same as b) except more conservative upper flux limits ($2\sigma + F$)
were adopted, see text for details.
}
\label{fig:N-S850-S1200}
\end{center}
\end{figure}

\begin{figure} 
\begin{center} 
\includegraphics[width=1.02\hsize,angle=0]{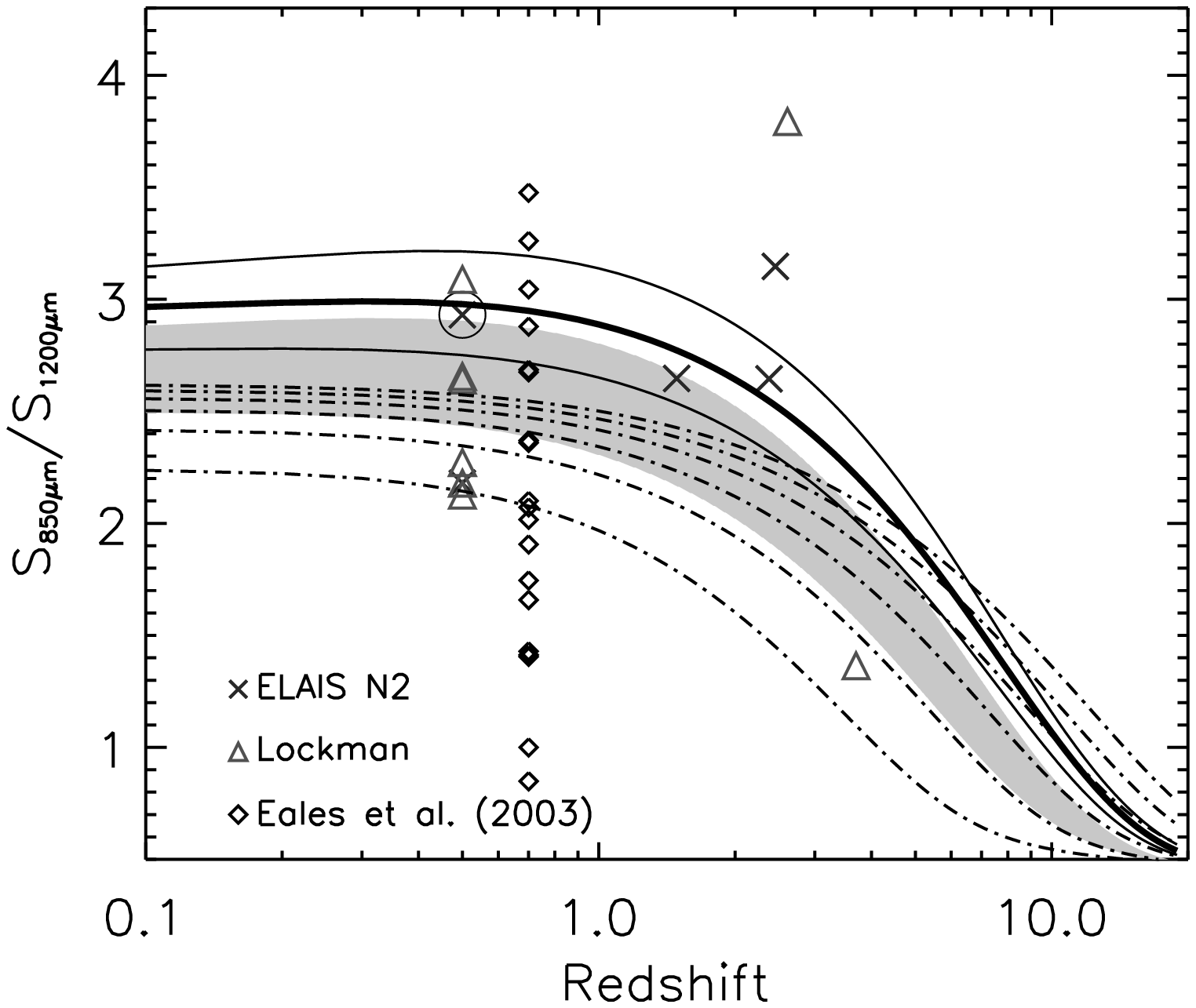}
\caption[$850/1200$\mum flux density ratio for SCUBA sources
detected by MAMBO.]  { 
The $850/1200$\mum flux density ratio
for SCUBA sources detected by MAMBO in the ELAIS\,N2 (crosses) and
Lockman Hole (triangles). Six of the sources have spectroscopic
redshifts (see Chapman et al.~2003, 2004), while the remainder have
been placed at $z\rm =0.5$. The two encircled sources have no radio
counterpart. Shown as diamonds are the $850/1200$\mum flux ratios for
the sample of Eales et al.~(2003). The thick solid curve represents
the expected flux ratio for an optically thin modified blackbody with
$T_{\rm d} = 45$\,K and $\beta=+1.5$, typical for a local ULIRG. This
curve is enveloped by two thin solid curves which correspond to
changes in the dust temperature and spectral index of $\Delta T_{\rm
d} = \pm 5$\,K and $\Delta \beta = \pm 0.2$. The light-grey shaded
area corresponds to a modified blackbody with $T_{\rm d} = (35.6 \pm
5.0)$\,K and $\beta=(+1.3\pm 0.2)$.  The dot-dashed curves corresponds
to SEDs with $\beta=1.0$ and $T_d=20 ... 70$\,K in step if 10\,K.}
\label{fig:z-S850-S1200}
\end{center}
\end{figure}

An additional 24\,hr of $850$\mum data --- not used in the original 8\,mJy
Survey --- resulted in a slightly different source catalogue than that given in
Scott et al.~(2002).  With the inclusion of the new data, the significance of
N2\,850.17 drops from $3.5\sigma$ to $3.3\sigma$, with a new flux density of
$5.3\pm 1.7$\,mJy, while N2\,850.16 disappears (see Ivison et al.~2002).  At
the position of N2\,850.17 in the MAMBO map, we detect a source at the
$2.7\sigma$ significance level, suggesting that N2\,850.17 is a real source,
even though it is just below the formal $3.5\sigma$ detection threshold of the
8\,mJy Survey.  N2\,850.16, however, is not detected at $1200$\mum at greater
than $1\sigma$ significance, confirming that this was a spurious source in the
original Scott et al.~map.

Based on extremely deep radio imaging of the 8\,mJy fields, Ivison et
al.~(2002) concluded that six of the original SCUBA sources in the ELAIS\,N2
and Lockman Hole fields were likely to be fake sources.  Not only were they
extracted from extremely noisy regions, but they also lacked a radio
counterpart despite being amongst the brightest SMGs in the sample. The sources
in question were LE\,850.9, LE\,850.10, LE\,850.11, LE\,850.15, LE\,850.20, and
N2\,850.14. From Figure \ref{fig:mambo_scuba_overlay_sources} it is seen that
none of these sources, which are denoted by grey crosses, coincide with a
$\ge$$3.5\sigma$ MAMBO source.  In fact, the highest significance 1200\mum
detection of the above sources was at the 1.6-$\sigma$ level. Hence, our MAMBO
maps strengthen the conclusion of Ivison et al.~(2002) that these sources are
spurious.  If we compare the MAMBO catalogue with the refined SCUBA source
catalogue, in which the above mentioned SCUBA sources (including N2\,850.16 and
N2\,850.17) have been omitted, we find that at the $\ge$$3.0/3.5/4.0\sigma$
level, we confirm 6/4/3 out of 33/14/7 SCUBA sources in the ELAIS\,N2 field and
9/8/5 out of 31/16/9 SCUBA sources in the Lockman Hole, respectively. Hence,
the MAMBO identification rate of SCUBA sources in the Lockman Hole increases to
56 per cent, which is becoming comparable to fraction of SMGs detected in deep
radio maps (Smail et al.~2000; Ivison et al.~2002).

An important point to make in this context is that Figure
\ref{fig:mambo_scuba_overlay_sources} clearly shows that while not all
SCUBA sources are detected by MAMBO, there is a good overall spatial
correspondence between SCUBA and MAMBO sources, as was also pointed
out in section \ref{section:clustering}. One way to interpret this
result is that the two surveys represent two different realisations of
the same large-scale structure. From two independent but similar
(sub)mm surveys of the same region, we would expect to find the most
significant sources in both surveys, but this is not necessarily true of
the fainter sources near the detection threshold.

%
% Table 4 
%
\begin{table*}
%\scriptsize
\caption{MAMBO and VLA radio identifications of the SCUBA 8\,mJy Survey
$\ge$$3.5\sigma$ source catalogue (Scott et al.~2002). The $850$\mum
and 1.4\,GHz radio flux densities are from Scott et al.~(2002) and
Ivison et al.~(2002), respectively.}
\vspace{0.5cm}
\begin{center}
\begin{tabular}{lclcccc}
\hline
\hline
SCUBA ID    &   $S_{850\mu\rm{m}}\pm \sigma_{850\mu\rm{m}}$ & MAMBO ID     &  $S_{1200\mu\rm{m}} \pm \sigma_{1200\mu\rm{m}}$   &   SCUBA/MAMBO Offset & RADIO   &  $S_{1.4\rm{GHz}}\pm \sigma_{1.4\rm{GHz}}$\\
            &              mJy                                  &              &   mJy                           &    $\arcsecs$       &         &   $\mu$Jy      \\ \hline
\multicolumn{2}{c}{$ \sigma \geq 4.0$ Detections} \\ 
\hline
N2\,850.1   &   $11.2\pm 1.6$                                   &              &                                 &    &    yes    &  $45\pm 16$\\
N2\,850.2   &   $10.7\pm 2.0$                                   & N2\,1200.39  &   $3.4\pm 1.1$                & 8.4&    yes    &  $92\pm 16$\\
N2\,850.3   &   $8.5\pm 1.6$                                    & N2\,1200.19  &   $2.9\pm 0.8$                & 5.1&    no     &  $<44$\\
N2\,850.4   &   $8.2\pm 1.7$                                    & N2\,1200.10  &   $3.1\pm 0.7$                & 4.5&    yes    &  $221\pm 17$\\
N2\,850.5   &   $8.5\pm 2.2$                                    & N2\,1200.3   &   $3.9\pm 0.8$                & 1.0&    yes    &  $77\pm 31$\\
N2\,850.6   &   $9.2\pm 2.4$                                    &              &                                 &    &    no     &  $38\pm 19$\\
N2\,850.7   &   $9.0\pm 2.4$                                    & N2\,1200.4   &   $3.4\pm 0.7$                & 4.0&    yes    &  $159\pm 27$\\
\hline
\multicolumn{2}{c}{$4.0 > \sigma \geq 3.5$ Detections} \\ 
\hline
N2\,850.8   &   $5.1\pm 1.4$                                    &              &                                 & &    yes    &  $74\pm 29$\\
N2\,850.9   &   $9.0\pm 2.5$                                    &              &                                 & &    yes    &  $33\pm 12$\\
N2\,850.10  &   $5.4\pm 1.5$                                    &              &                                 & &    no     &  $58\pm 24$\\
N2\,850.11  &   $7.1\pm 2.0$                                    &              &                                 & &    no     &  $<44$\\
N2\,850.12  &   $5.5\pm 1.6$                                    &              &                                 & &    no     &  $32\pm 17$\\
N2\,850.13$^\ddag$  &   $6.3\pm 1.9$                                    &              &                                 & &    yes    &  $99\pm 23$\\
N2\,850.14$^*$  &   $11.2\pm 3.3$                               &              &                                 & &    no     &  $<44$  \\
N2\,850.15  &   $5.0\pm 1.5$                                    &              &                                 & &    no     &  $31\pm 20$\\
N2\,850.16$^\dagger$  &   $12.9\pm 3.9$                                   &              &                        &          &    no     & $<44$ \\
N2\,850.17$^{\dagger \dagger}$  &   $5.3\pm 1.7$                                    &              &               &                   &    no    & $<44$ \\
\hline
\hline
\multicolumn{2}{c}{$ \sigma \geq 4.0$ Detections} \\ 
\hline
LE\,850.1   &   $10.5\pm 1.6$                                   & LE\,1200.5   &   $3.4\pm 0.6$                  & 5.1 &    yes    &  $73\pm 10$\\
LE\,850.2   &   $10.9\pm 2.4$                                   & LE\,1200.1   &   $4.8\pm 0.6$                  & 1.3&    yes    &  $29\pm 11$\\
LE\,850.3   &   $7.7\pm 1.7$                                    & LE\,1200.11  &   $2.9\pm 0.7$                  & 6.2&    yes    &  $98\pm 12$\\
LE\,850.4   &   $8.3\pm 1.8$                                    &              &                                 & &    no     &  $19\pm 8$\\
LE\,850.5   &   $8.6\pm 2.0$                                    &              &                                 & &    no     &  $<25$\\
LE\,850.6   &   $11.0\pm 2.6$                                   & LE\,1200.10  &  $2.9\pm 0.7$                   & 7.7&    yes    &  $54\pm 14$\\
LE\,850.7   &   $8.1\pm 1.9$                                    &              &                                 & &    yes    &  $135\pm 13$\\
LE\,850.8   &   $5.1\pm 1.3$                                    & LE\,1200.14  &  $2.4\pm 0.6$                   & 3.9&    yes    &  $58\pm 12$\\
LE\,850.9$^*$   &   $12.6\pm 3.2$                               &              &                                 & &    no    &  $<23$\\
LE\,850.10$^*$  &   $12.2\pm 3.1$                               &              &                                 & &    no    &  $<25$\\
LE\,850.11$^*$  &   $13.5\pm 3.5$                               &              &                                 & &    no    &  $26\pm 12$\\
LE\,850.12  &   $6.2\pm 1.6$                                    &              &                                 & &    yes    &  $278\pm 12$\\
\hline
\multicolumn{2}{c}{$4.0 > \sigma \geq 3.5$ Detections} \\ 
\hline
LE\,850.13  &   $9.8\pm 2.8$                                    &              &                                 & &    no     &  $18\pm 11$\\
LE\,850.14  &   $9.5\pm 2.8$                                    & LE\,1200.3   &  $3.6\pm 0.6$                   & 2.0&    yes    &  $72\pm 12$\\
LE\,850.15$^*$  &   $11.7\pm 3.4$                               &              &                                 & &    no    &  $<21$\\
LE\,850.16  &   $6.1\pm 1.8$                                    & LE\,1200.6   &  $2.8\pm 0.5$                   & 2.6&    yes    &  $41\pm 12$\\
LE\,850.17  &   $9.2\pm 2.7$                                    &              &                                 & &    no     &  $<23$\\
LE\,850.18  &   $4.5\pm 1.3$                                    & LE\,1200.12  &  $3.3\pm 0.8$                   & 3.1&    yes    &  $47\pm 10$\\
LE\,850.19  &   $5.5\pm 1.6$                                    &              &                                 & &    no     &  $<27$\\
LE\,850.20$^*$  &   $10.3\pm 3.1$                               &              &                                 & &    no    &  $<24$\\
LE\,850.21  &   $4.5\pm 1.3$                                    &              &                                 & &    yes    &  $21\pm 10$\\
\hline
\multicolumn{5}{l}{$^\ddag$ Detected with MAMBO at the $2.8\sigma$ level.} \\ 
\multicolumn{5}{l}{$^*$ Excluded from the refined 8\,mJy sample of Ivison et al.~(2002).} \\ 
\multicolumn{5}{l}{$^\dagger$ This source vanished with the inclusion of an additional 24\,hr of SCUBA data.} \\ 
\multicolumn{5}{l}{$^{\dagger \dagger}$ This source dropped from $3.5$ to $3.3\sigma$ with the inclusion of additional SCUBA data. It is
detected with MAMBO at $2.7\sigma$.} \\ 
\label{table:scubo-list}
\end{tabular}
\end{center}
\end{table*}

Deep radio observations provide an alternative route to reliably
identify SMGs. About two thirds of SMGs are detected in the radio
where the ratio of submm to radio flux detection thresholds is above
$\sim$400 (Smail et al.~2000; Ivison et al.~2002).  However, it
remains an open question whether the third of the population which are
radio-blank are SMGs at very high redshifts ($z >> 3$), or cooler,
less-far-IR-luminous objects at similar redshifts as the bulk of the
population, or simply spurious sources. All three scenarios would
explain the lack of radio counterparts.

In Table \ref{table:scubo-list} we list all the $\ge$$3.5\sigma$ SCUBA
sources with a MAMBO counterpart detected at $\ge$$3.0\sigma$
significance within 10$\arcsecs$, along with their positional offsets
and flux densities at $1200$ and $850$\mumeol. Of the 12 $\ge$$3.5\sigma$
SCUBA sources in the ELAIS\,N2 field which were not detected by MAMBO,
only four had a radio identification.  Of the eight radio-blank SCUBA
sources, only one --- N2\,850.3 --- was confirmed by MAMBO, indicating
that it could be cool, or lie at $z >> 3$.  In the Lockman Hole, {\em
none} of the radio-blank SCUBA sources were detected at $1200$\mumeol,
and only three of the 11 radio-identified SCUBA sources were not detected
by MAMBO.

Since the depth of the MAMBO maps at $1200$\mum is comparable to that of
the SCUBA maps at $850$\mum it is hard to conceive of a way in which an
$850$\mum source with no radio counterpart could fail to be detected
at $1200$\mum. The only plausible explanations require that the
sources are spurious or confused.
 
In Table \ref{table:scubo-list-summary} we have summarised the above
findings.  They suggest strongly that the fraction of robust SCUBA
sources, i.e.~sources confirmed by MAMBO, which are not detected in
the radio is low: 20 per cent (1/5) in ELAIS\,N2 and 0 per cent
in the Lockman Hole.  These findings suffer from small number
statistics and we await a comparison between the much larger map of
the Lockman Hole being obtained with MAMBO (Bertoldi et al.~in prep.)
and the SCUBA map of that region which is being obtained as part of
SHADES (see {\it http://www.roe.ac.uk/ifa/shades/}).  

In the light of current and future (sub)mm surveys, our findings
underline the importance of multi-wavelength follow-up in order to
establish the reality of (sub)mm sources, and the dependence only on
the most robust samples to draw meaningful statistical conclusions
(cf.~Dannerbauer et al.~2004).

%
% Table 5
%
\begin{table}
\scriptsize
\caption{The number (and percentage) of SCUBA sources identified by
MAMBO divided into radio and non-radio ID categories. Note, that all
the number in this table are based on the comparison with the
$\ge$$3.5\sigma$ 8\,mJy Sample of Scott et al.~(2002) and the
$\ge$$3.0\sigma$ MAMBO sample presented in this paper.}
\vspace{0.2cm}
\begin{center}
\begin{tabular}{lcccccc}
       &\multispan3{\hfil MAMBO ID \hfil}&\multispan3{\hfil No MAMBO ID  \hfil} \\
Field  &&&& \\
        &Radio ID & No Radio ID &Total&Radio ID &No Radio ID&Total    \\
\noalign{\smallskip}
ELAIS\,N2     &4            &1          &5	&4 	        &8	        &12\\
Lockman       &8            &0          &8	&3 	        &10	        &13\\
\noalign{\smallskip}
Total         &12 (92\%)    &1 (8\%)    &13	&7 (28\%)	&18 (72\%)	&25\\
\label{table:scubo-list-summary}
\end{tabular}
\end{center}
\label{table:scubo-list-summary}
\end{table}

\subsection{The $850/1200$\mum flux density ratio}

Another valuable piece of information which can be gleaned from a
comparison of the MAMBO and SCUBA maps is the $850/1200$\mum flux ratios
for a large sample of (sub)mm galaxies.  Beyond $z \geq\rm 3$ this flux
ratio becomes a strong function of redshift, and can thus be used as a
crude discriminator between low- and high-redshift sources (Eales et
al.~2003), much in the same way that the radio-to-submm spectral
index acts as a redshift estimator for sources at $z\ls\rm 3$ (Carilli \&
Yun 1999, 2000).  Thus, the two redshift estimators complement
each other, and if used in combination can potentially be used to
probe the redshift distribution of (sub)mm sources.  However, both of
these redshift estimators suffer from the $T_{\rm d}-z$ degeneracy first
pointed out by Blain (1999), which implies that a low-redshift
source with a cold dust temperature is indistinguishable from a warm
source at high redshift.

The distribution of $850/1200$\mum flux ratios for the 13 $\ge$$3.5\sigma$
SCUBA sources which were robustly identified by our MAMBO survey are plotted in Figure
\ref{fig:N-S850-S1200}a, along with the ratios for a sample of SCUBA-observed
MAMBO sources by Eales et al.~(2003).  Using SCUBA in its photometry mode, they
observed 21 MAMBO-selected sources from the MAMBO Deep Field Survey 
(Bertoldi et al., in preparation).  While there is a considerable overlap
between the two distributions, the latter has a significant over-density at low
values which is not reproduced by our sample. From our sample we find a median
value of $S_{850}/S_{1200} = 2.6\pm 0.6$, marginally higher than the median
value of $2.1\pm 0.7$ found by Eales et al.~(2003), although within the
scatter.

In Figure \ref{fig:z-S850-S1200} we have plotted the measured $850/1200$\mum
flux ratios of our sample and the Eales et al.~sample along with curves showing
the predicted $850/1200$\mum flux density ratios against redshift for an SED
with $T_{\rm d} = 45$\,K and $\beta=+1.5$, and an SED with $T_{\rm d} =
35.6$\,K and $\beta=+1.3$. The former SED provides a good match to the local
ULIRG, Arp\,220, while the latter is based on the average $T_{\rm d}$ and
$\beta$ values derived from the SCUBA Local Universe Galaxy Survey (Dunne et
al.~2000). In addition, we have also plotted the
$S_{850\mu\rm{m}}/S_{1200\mu\rm{m}}$--$z$ relationship for a set of SEDs with
$\beta=1$ and $T_d=20,30 ... 70$\,K, i.e.\ SEDs which have very different dust
properties from local ULIRGs. Note, the above dust temperatures are at a
redshift of zero.  In order to account for the increasing cosmic microwave
background radiation (CMBR) temperature with redshift and its thermal coupling
with the dust we have used
\begin{equation}
T_d(z) = \left ( T_{d,z=0}^{4+\beta} + T_{\rm CMBR}^{4+\beta} \left [ (1 + z)^{4+\beta} - 1 \right ] \right ) ^{1/(4+\beta)},
\end{equation}
where $T_{\rm CMBR}$ is the present epoch CMBR temperature, (see e.g.~Eales \&
Edmunds 1996).  The effect is only significant when the dust is cold relative
to the CMBR temperature, and therefore becomes more important at very high
redshifts. The slight downward trend in the
$S_{850\mu\rm{m}}/S_{1200\mu\rm{m}}$ curves at low redshifts is due to the
negative spectral slope ($\alpha = -0.7$) of the radio continuum emission which
boosts the $1200$\mum flux relative to the flux at $850$\mum. However, this
effect is completely negligible at redshifts beyond $z\sim\rm 0.5$.

The low flux ratios found by Eales et al.~(2003) led them to conclude that a
significant fraction of SMGs must lie at very high redshifts ($z\rm >> 3$) or
possess dust properties different from low-redshift starburst galaxies.  In
contrast, we find no conflict between our measured
$S_{850\mu\rm{m}}/S_{1200\mu\rm{m}}$ ratios and SEDs based on local
ULIRGs. From a subset of five sources which have been targeted
spectroscopically by Chapman et al.~(2003, 2004), and thus are placed at their
correct redshift in Figure \ref{fig:z-S850-S1200}, we can conclude that the
observed flux ratios for at least four of the five sources are consistent with
the range of SEDs expected from local ULIRGs.  The one exception is the
outlying point at $S_{850\mu\rm{m}}/S_{1200\mu\rm{m}}=\rm 3.8$ This data point
corresponds to the source LE\,850.6/LE\,1200.10 which has another source
(LE\,1200.9) close to it. This source was detected by MAMBO but not by SCUBA,
and it is likely that LE\,1200.9 contributes to the 850\mum flux of
LE\,850.6, leading to an artificially high flux ratio for this source.

Our sample is taken from two unbiased surveys at slightly different
(sub)mm wavelengths and is thus independent of any radio selection
bias.  Furthermore, it is clear from Figure \ref{fig:z-S850-S1200}
that the SMG without a radio counterpart (circled symbol) does not
have a lower $850/1200$\mum flux ratio than the rest of the sample. This
suggest that the source is blank in the radio because it is a
cooler, less-far-IR-luminous object at similar redshifts to the bulk
of the population, not because it lies at a very high redshift.

Candidates for very-high-redshift sources, i.e.~850\mum dropouts, should be
sought amongst sources detected by MAMBO but not SCUBA.  From Figure
\ref{fig:mambo_scuba_overlay_sources} it is seen that 9 MAMBO sources in the
ELAIS\,N2 field and 9 in the Lockman Hole fall within the regions observed by
SCUBA, yet are not detected at $\ge$$3.5\sigma$ significance at
$850$\mumeol. Upper limits on the flux densities at $850$\mum of these sources were measured, taking the
peak flux in a 10\arcsec radius aperture region of the SCUBA map coincident with
the MAMBO position.  These limits were then merged with the $850$\mum
fluxes of the robust sample, resulting in $850$\mum flux
estimates (or upper limits) for all the 9+9+13=31 MAMBO
sources which lie within the SCUBA regions.
The resulting distribution of $850$-to-$1200$\mum flux ratios 
is shown as the dotted curve in Figure
\ref{fig:N-S850-S1200}b. 
The distribution appears to have two peaks, one at
$S_{850\mu\rm{m}}/S_{1200\mu\rm{m}}\sim 1$ which reproduces
the low-end tail of Eales et al.~(2003) rather well and 
is almost entirely due to the MAMBO sources not detected by SCUBA, and another at $\sim
2.5$ which stems from the 13 sources robustly identified at both
1200\mum and 850\mumeol. 

In order to make a fair comparison with the flux ratios of the
SCUBA sample, we estimated 1200\mum fluxes for all the
SCUBA sources. This was done in an identical fashion as above, i.e.
upper 1200\mum flux limits were derived for the SCUBA sources not detected with MAMBO 
using the peak flux within a 10\arcsecs radius aperture region of the MAMBO map
centered on the SCUBA position, and concatenated with the robust sample.
This distribution is shown as the dashed histogram in \ref{fig:N-S850-S1200}b.

The dotted and dashed distributions appear to be distinct, and
in order to determine the probability of the two samples being
drawn from the same parent distribution, we have employed
the standard 'survival analysis' tests (Feigelson \& Nelson 1985),
which are appropriate in the case where the samples contain upper or lower limits
(censored data). This included the Gehan, log rank, Peto-Peto, and Peto-Prentice tests, the
latter being perhaps the most conservative and least sensitive to differences
in the censoring patterns. 
However, common to all the tests is that they are unable to compare
samples with mixed censor indicators, i.e. one cannot compare
a sample containing upper limits with a sample containing lower limits. 
As a result we ran the tests where only one of the samples contained 
censored data, assuming that the limits in the other sample were
not limits but measured values, and vice versa. 
In both cases the tests yielded probabilities
less than 0.01, thereby strongly suggesting that 
the two distributions are significantly different.
This argues in favour of there being a low-end tail
consisting of galaxies at either high redshifts
or with cool dust temperatures.
Consistent with these findings is the
distribution of MAMBO sources with radio counterparts
(shaded histogram in Figure \ref{fig:N-S850-S1200}b)
which shows that few of the sources with flux ratios
$\ls 1.5$ are identified in the radio. This
is what one would expect if they were at 
high redshifts or cool, not very far-IR luminous
systems.

In addition to the two astrophysical explanations (i.e.~cold dust or
high redshift) Eales et al.~also suggest more mundane reasons for
their low flux ratios, the most important of which is astrometric
errors (see discussion in Eales et al.~2003).  For all but five of
their 21 sources, they used positions derived from radio observations
or mm-wave interferometry. For the remaining sources they used the
MAMBO positions.  With our dataset, in conjunction with the 8\,mJy
Survey images, we can reproduce the Eales et al.~experiment in the case
where only MAMBO positions are available. In order to do this we consider
the sample of 13 sources which were detected by both MAMBO and SCUBA. However,
instead of using the 850\mum fluxes reported by Scott et al.~(2002), we 
measure the 850\mum flux at the MAMBO position in the SCUBA
map using the peak flux within a 10\arcsecs aperture. 
In general, we find that the effects of positional errors are small, 
and the flux ratios are not skewed towards lower
values.

Finally, it is possible that contamination by spurious or flux-boosted MAMBO
sources is responsible for at least some of the low flux ratios observed.
However, based the Monte Carlo simulations in section
\ref{section:monte-carlos} we expect no more than 2 sources to be spurious in
each of the MAMBO maps. This is an upper limit, given that the overlap between
the SCUBA and MAMBO regions is only about 68 per cent of the areas covered by
MAMBO.

While the above analysis suggests that astrometrical errors and spurious/flux-boosted
sources are unable to account for the 12 MAMBO sources with $S_{850\mu\rm{m}}/S_{1200\mu\rm{m}}\ls \rm 1.5$,
we caution that the low flux ratio could be due to the way we have estimated
the upper flux limits. Simply adopting the peak flux
within a 10\arcsec aperture might in some cases result
in too low flux estimates and would tend to bias the
MAMBO and SCUBA distributions towards lower and higher 
flux ratios, respectively. In Figure \ref{fig:N-S850-S1200}c
we have adopted a more conservative approach in which
the upper limits on the fluxes were estimated by
adding $2\times \sigma$ to the peak flux, where
$\sigma$ is the local rms noise. Clearly, the overlap 
between the two distributions is now much greater, and
the two distributions appear to be indistinguishable. 
This is confirmed by the 'survival
analysis' tests which yield probabilities in the range
0.22 (log rank) to 0.11 (Peto-Prentice) and similar for the reverse comparison, i.e.
comparing the SCUBA distribution with the uncensored
MAMBO distribution. Thus, adopting what is arguably more
realistic upper flux limits we find no evidence for
SMGs with unusually low 850-to-1200\mum flux ratios as
reported by Eales et al.~(2003).
If this is the case, the major
implication is that, beyond the few completely
unrepresentative submm-loud AGN at $z>4$, there is no significant
population of SMGs at very high redshifts. The redshift
distribution of radio-identified SCUBA sources, as determined by
Chapman et al.~(2003, 2004), would be applicable to virtually all
of the (sub)mm population.

However, while we find no conclusive evidence for '850\mum-dropouts' a handful
of MAMBO sources do seem to be good candidates for SMGs at $z>>5$.
In particular, LE\,1200.2 is one of the brightest
sources in our survey, yet it is not detected by SCUBA nor is it seen in the
radio. Pointed SCUBA photometry observations of this source 
would provide an accurate accurate estimate of its 850\mum flux and confirm
or dismiss its status as a '850\mum-dropout'.

Furthermore, with the
completion of the MAMBO Deep Field Survey and the SCUBA Half Degree
Extragalactic Survey, which will not only provide us with much larger samples
but also a multitude of observations at complementary
wavelengths, we should be able to obtain a much better
census of the high-redshift tail of (sub)mm sources.

\section{Conclusions}

In this paper we have presented results from a MAMBO $1200$\mum
blank-field survey of the ELAIS\,N2 and Lockman Hole fields, covering
a total of 357 arcmin$^2$ to a rms level of
$\sim$0.8\,mJy\,beam$^{-1}$. We detect 27 sources at 
$\ge$$4.0\sigma$ significance, and more than 40 sources at
$\ge$$3.5\sigma$.

From the $\ge$$3.5\sigma$ catalogue we have derived accurate number
counts over the flux range 3--5.5\,mJy, and find evidence for a break
at $S_{1200\mu\rm{m}}\rm \simeq 4$\,mJy. This corresponds to a far-IR
luminosity of $\sim10^{13}\,\Lsolar$ for a modified blackbody with
$T_{\rm d}=40$\,K and $\beta = +1.5$ at $z\rm =2.5$. The observed
$1200$\mum source counts can be successfully reproduced by a simple
parametric model for the evolution the local ULIRG population.
Furthermore, this model also fits the $850$\mum source counts which
suggests that the MAMBO and SCUBA sources are drawn from the
same population of dust-enshrouded starburst at high redshift.

Two independent tests were carried out with the aim of detecting
clustering in the MAMBO population. Although, the angular two-point 
correlation function showed no evidence of clustering, a nearest 
neighbour analysis suggests that the most significant MAMBO
sources are not randomly distributed but come in pairs, typically
separated by 23$\arcsecs$. Furthermore, 
the spatial distribution of sources appears to be non-random, with sources
tending to reside in clusters surrounding large voids.
The reality of these structures is strengthened by the good overall
spatial correlation between the SCUBA, MAMBO and $\mu$Jy-level radio
sources. This suggests that co-spatial surveys at the two slightly
different (sub)mm wavelengths skim the brightest members of a numerous
but faint population, yielding two similar but low-signal-to-noise
visualisations of the true (sub)mm sky.

Our MAMBO survey confirms roughly half of the refined
$\geq$$3.5\sigma$ SCUBA 8\,mJy Survey sample (Scott et al.~2002;
Ivison et al.~2002).  This is comparable to the radio identification
rate of SCUBA and MAMBO sources. As a by-product of this analysis, we
have produced a extremely robust sub-sample of 13 SMGs detected at
$\ge$$3.5\sigma$ by both SCUBA and MAMBO. We find that only one
($\sim$8 per cent) has no radio counterpart, a significantly lower
fraction than the third which the radio-blank SMGs have generally been
believed to constitute. Our results thus suggest that the 
population of SMGs which are detected by both SCUBA and MAMBO has
no significant tail at $z\rm >> 3$.  This
conclusion is further strengthened by the observed distribution of
$850/1200$\mum flux density ratios for the 13 sources in our sample.  We
find their flux density ratios to be consistent with the SEDs found
for local ULIRGs and in agreement with the spectroscopic redshift
distribution of SMGs as determined by Chapman et al.~(2003, 2004).

Finally, we have identified 18 MAMBO sources within the SCUBA UK 8\,-mJy regions
which are not detected at $850$\mum at greater than $3.0\sigma$ significance. 
Any high-redshift SMGs should be sought amongst this population of 
'850\mum-dropouts'. However, using conservative upper flux limits 
we find that the distribution of $850$-to-$1200$\mum 
flux ratios for these sources is statistically indistinguishable
from that of the sources identified robustly in both wavelengths.

\section*{Acknowledgements}

TRG acknowledges support from the Danish Research Council and from the
European Union RTN Network, POE.  We are grateful to Robert Zylka for providing us
with the {\sc mopsi} package and for useful
discussions concerning {\sc mopsi}. We are also grateful to Ian Smail
for helpful comments on the paper. Finally, we thank Ernst Kreysa and his
team for providing MAMBO.

\bsp

%\label{lastpage}

\end{document}